\documentclass[final,1p,times]{amsart}
\usepackage{geometry}             
\usepackage{graphicx}
\usepackage{amssymb}
\usepackage{epstopdf}
\usepackage{xargs}                 
\usepackage[pdftex,dvipsnames]{xcolor}  
\usepackage[colorinlistoftodos,prependcaption,textsize=tiny]{todonotes}
\usepackage{siunitx}
\usepackage{graphicx}
\usepackage{caption}
\usepackage{subcaption}
\usepackage{verbatim}
\usepackage{verbatimbox}
\usepackage{xcolor, colortbl}
\usepackage[hidelinks]{hyperref}

\newcommandx{\unsure}[2][1=]{\todo[linecolor=red,backgroundcolor=red!25,bordercolor=red,#1]{#2}}
\newcommandx{\change}[2][1=]{\todo[linecolor=blue,backgroundcolor=blue!25,bordercolor=blue,#1]{#2}}
\newcommandx{\info}[2][1=]{\todo[linecolor=OliveGreen,backgroundcolor=OliveGreen!25,bordercolor=OliveGreen,#1]{#2}}
\newcommandx{\improvement}[2][1=]{\todo[linecolor=Plum,backgroundcolor=Plum!25,bordercolor=Plum,#1]{#2}}
\newcommandx{\thiswillnotshow}[2][1=]{\todo[disable,#1]{#2}}

\newcommand{\sbwi}{single-beam Wollaston interferometer}
\newcommand{\mlem}{Maximum Likelihood -- Expectation Maximisation}

\title{Realtime Tomography of Gas-Jets with a Wollaston Interferometer}
\author{A.\ Adelmann, B.\ Hermann, R.\ Ischebeck, M.C.\ Kaluza, U.\ Locans, N. Sauerwein \& R. Tarkeshian}
\date{\today}                                           

\begin{document}
\begin{abstract}

A tomographic gas-density diagnostic using a \sbwi\  able to characterise  
non-symmetric density distributions in gas jets is presented.\ A real-time tomographic algorithm is able to reconstruct three dimensional density distributions. A \mlem\ algorithm, an iterative method with good convergence properties compared to simple back projection, is used.\ With the use of graphical processing units, real time computation and high resolution are achieved.\ Two different  gas jets 
are characterised: a kHz, piezo-driven jet for lower densities and a solenoid valve based  jet producing higher densities. 
While the first jet is used for FEL photon beam characterization, the second jet is used in laser wakefield acceleration experiments.\ In this latter application, well-tailored and non-symmetric density distributions produced by a supersonic shock front generated by a razor blade inserted laterally to the gas flow, which breaks cylindrical symmetry, need to be characterized.\

\end{abstract}

\maketitle
\section{Introduction}
The aim of future advanced accelerator concepts is to use accelerating structures that can sustain higher electric field strengths to downsize the structure compared to conventional microwave cavities.\ 
Laser wakefield acceleration (LWFA) is one such promising technology, which is being investigated by several research groups around the world.\ The idea of LWFA is to excite a wake field in an underdense plasma by a short energetic laser pulse, the so-called driver.\ The plasma can be generated by ionising a gas jet using the driver pulse itself.\ In this plasma, the driver pulse excites electron density oscillations, which are co-propagating with the driver almost at the speed of light forming a plasma wave, which is moving with a relativistic phase velocity. Due to the local charge separation---the ions remain stationary on the corresponding time scales---longitudinal electric fields with amplitudes in excess of 100\,GV/m are generated, orders of magnitude higher than in conventional accelerator structures. Electrons can be accelerated in this moving field structure when they are trapped in the wave with the right phase, i.\ e.\ at the correct position and with correct velocity.\ An inherently synchronized method for trapping electrons is down-ramp injection, which can be induced when the plasma wave propagates across a sudden drop in density, a so-called down-ramp \cite{PhysRevE.58.R5257,PhysRevLett.100.215004}.\ Here, the plasma wave, which has been generated in a first, high-density region, enters a second region with lower density.\ The electrons in the high-density region oscillate faster since the plasma frequency  $\omega_{p} = \sqrt{n_{e}e^{2}/(\varepsilon_0 m_e)}$ depends on the electron density $n_{e}$; $m_e$ and $e$ are mass and charge of the electron, respectively.\ Therefore, the wavelength of the plasma wave, $\lambda_p\approx c/\omega_p$, is shorter than in the subsequent low-density region. When this density transition occurs over a distance of the order or shorter than the plasma wavelength, some of the electrons forming a density peak in the plasma wave in the high-density region will then no longer be associated with the density peak of the low-density plasma wave but they will find themselves in the accelerating phase of the wave.\ It has been shown that a density change by a factor of 2 to 3 can be achieved by a shock front in supersonic gas jets \cite{downramp}.\ At such a sharp transition, many electrons are loaded into the same phase-space volume of the plasma wave, which will result in a narrow energy spread of the accelerated electrons.\ The parameters of the generated electron beam (energy spectrum, charge and duration) critically depend on the plasma wave and its evolution, which in turn is very sensitive to the plasma density distribution. With these facts in mind, a precise and fast real-time density measurement is required for controlling the down ramp process \cite{PhysRevAccelBeams.20.051301}, and---when shot-to-shot-fluctuations in the gas jet cannot be sufficiently suppressed---for increasing the shot-to-shot stability of the generated electron beam.\ The view of high quality beams from LFWA, a fast and accurate characterization of the gas jet is of great importance. 
A comprehensive overview on density measurements of gas jets can be found in \cite{Eckbreth1996}.

While cylindrically symmetric gas jet distributions can easily be characterized from a single interferometric measurement taken in one direction, e.g. using a Michelson-type interferometer or a Nomarski-type interferometer using a Wollaston prism, tomographic methods are necessary when non-symmetric distributions, e.g.\ caused by a density jump as mentioned above, are required. The latter methods, which include interferometric measurements of the gas density taken along many directions, are more elaborate and sometimes time-consuming to analyse, when high accuracy and spatial resolution are needed \cite{PhysRevE.49.5628,PhysRevE.50.4266,PhysRevE.54.6769}. Therefore, analysis methods which can be applied in quasi-real time are highly advantageous. Here, we present the physics setup and the computational method, which fulfils above demands.

The paper is organised as follows: Section \ref{sec.wprism} introduces the  principle of the density measurement using a Wollaston prism.\ The measurement is then put in context to LWFA application.\ In section \ref{sec:analysis}, the details of the optical set-up as well as the set-up for tomography of a non-symmetric density distribution are introduced.\ Section \ref{sec:ExampleMeasurements} presents experimental results from the characterisation of different gas jets, while section \ref{sec:Conclusion} summarizes the paper.

\section{The Single Beam Wollaston Interferometer Set-Up} \label{sec.wprism}

\subsection{Theory} \label{sec:theory}
\begin{figure}[h!]
\begin{center}
\begin{subfigure}[b]{0.48\textwidth}
\includegraphics[width=\textwidth]{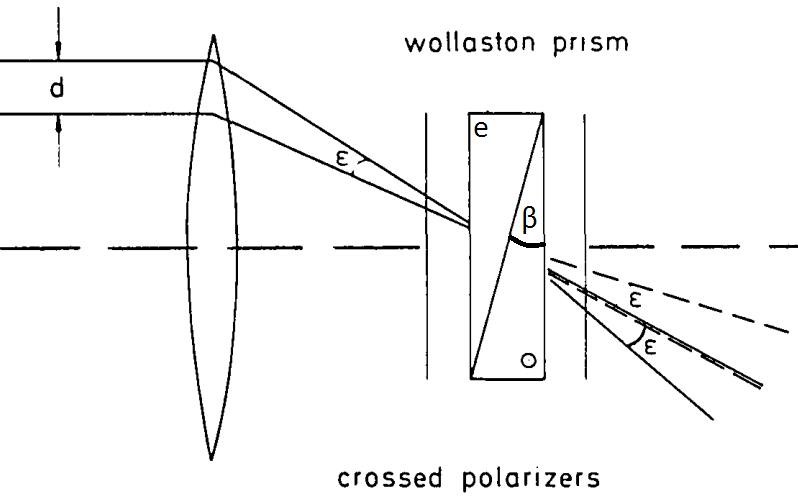}
\caption{Working principle of the Wollaston prism \cite{Merzkirch}.}
\label{fig.wprism}
\end{subfigure}
\hfill
\begin{subfigure}[b]{0.48\textwidth}
\includegraphics[width=\textwidth]{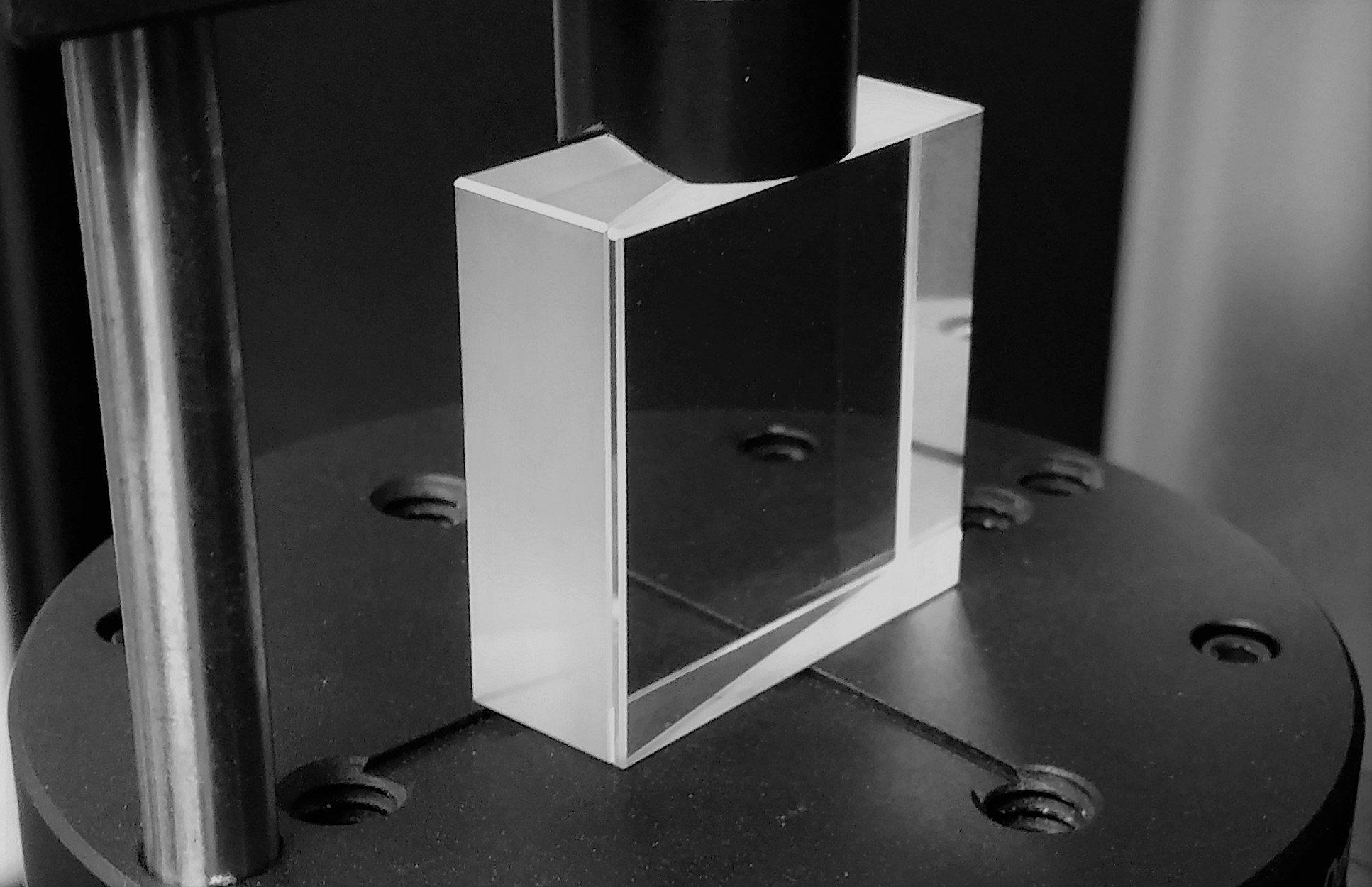}
\caption{Wollaston prism 30x30x5 mm, $\epsilon=5.8$ mrad, Societe d'Optique de Pr\'ecision Fichou.}
\label{fig.wprism2}
\end{subfigure}
\caption{Working principle of the Wollaston prism (A) and used prism (B).}
\end{center}
\end{figure}

A Wollaston prism (figure \ref{fig.wprism2}) is a combination of two prisms made of a birefringent crystal (e.g.\ quartz), which has a single optical axis.\ The two prisms are cut and combined such that the optical axes of the prisms are perpendicular to each other.\ The main feature of such a birefringent crystal is that the refractive index for light polarized parallel to the optical axis ($\eta_e$) is different than for light polarized perpendicular to the optical axis ($\eta_o$).\ Hence, a light ray polarized at $45^\circ$ with respect to the optical axis is split up in two equally intense beams with perpendicular polarization, which experience different refractive indices and hence propagate with different phase velocities.\ In a Wollaston prism, one of the two beams experiences a higher and the second a lower refractive index in the first birefringent prism, while the situation is reversed in the second prism. Hence, the two rays, which still propagate collinearly in the first prism will be refracted differently at the interface between the two prisms and then propagate under an angle in the second prism. After exiting the second prism at its back surface, the two rays will diverge with an angle $\epsilon$ (dispersion angle), which depends on the geometric and optical properties of the crystals forming the Wollaston prism.\ As shown by Small \cite{Small},
\begin{equation*}
   \epsilon=2 \beta (\eta_e-\eta_o) \text{,} 
\end{equation*}
where $\beta$ is the prism angle of both prisms (see figure\ \ref{fig.wprism}).\ The Wollaston prism used in this experiment was manufactured by Societe d'Optique de Pr\'ecision Fichou (figure \ref{fig.wprism2}) and has a dispersion angle of \SI{20}{\arcsecond} = 5.8 mrad.\ If two light rays with initial relative angle $\epsilon$ pass through the prism, one polarization component of the first ray will coincide with the polarization component of the second ray (compare figure\ \ref{fig.wprism}).\ Before passing through the lens these two rays are separated by a distance $d=\epsilon f_\circ$.\ For the Wollaston prism and a lens with focal length $f_\circ=30\,$cm, $d = 1.74$\,mm.\ If a second polarizer at $-45^\circ$ is placed after the prism, these two rays can interfere.\ This concept is illustrated in figure \ref{fig.wprism}.\ 
Depending on the position of the prism relative to the focal point of the lens the interfering rays will pick up a phase difference.\ If the prism is centred at the focal point, every optical path runs equal distances through the two halves of the prism, i.e.\ no relative phase difference is generated.\ This scenario is called normal mode or infinite fringe width (IFW) set-up.\ Displacing the prism by a distance $b$ (see figure \ref{fig.setup1}) from the focal point results in different path lengths of interfering rays, as they travel different distances in each half of the prism.\ These phase shifts lead to regular interference patterns on the screen.\ This set-up ($b\neq 0$) is called differential mode or finite fringe width (FFW) set-up \cite{Biss}.\ The FFW used in this work is sketched in figure \ref{fig.setup1}.\ The spacing $S$ between the undisturbed fringes is given by
\begin{equation}
    S = \frac{\lambda}{\epsilon}\frac{p}{b} \label{eq.fringe_width} \text{,} 
\end{equation}
where $p$ is the distance from the Wollaston prism to the screen \cite{bena}.\ 
The spacing can be decreased by increasing $b$, i.e. by changing the position of the Wollaston prism with respect to the focal plane.\ If the position of the screen remains fixed, which is the case when the lens images a certain plane onto the screen, $p$ changes accordingly.\ Placing a medium with refractive index $\eta\neq 1$ that covers only parts of the laser beam will result in a shift of the fringe spacing $S$, since rays passing through this medium will pick up an additional phase shift with respect to the unperturbed rays.\

\subsection{Estimation of Phase Shift due to a Gas Jet Density Distribution}\label{sec.est_dens}
\begin{figure}[h!]
\begin{center}
\includegraphics[width=9cm]{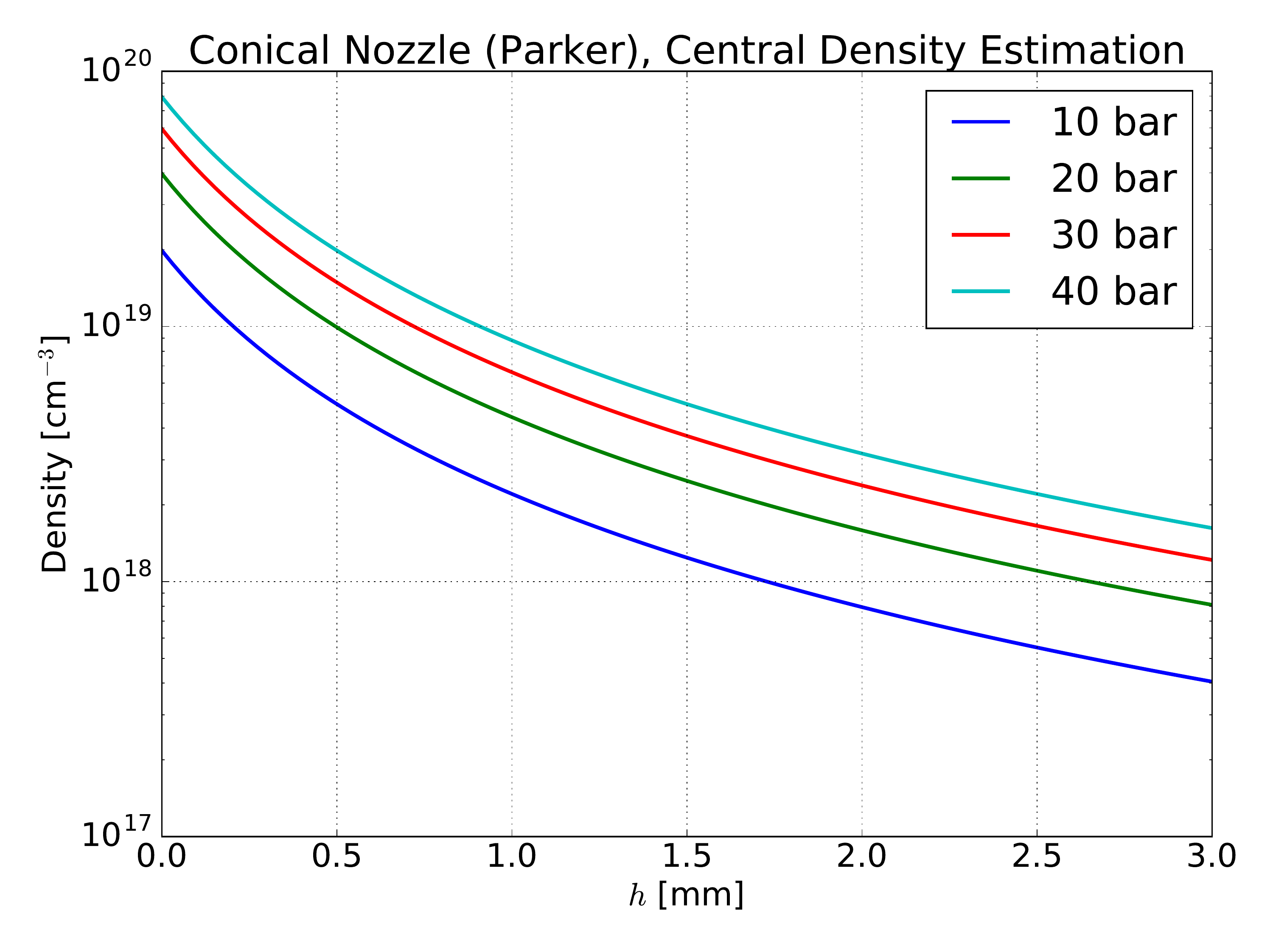}
\caption{Estimation of the on-axis density $\rho$ for different backing pressures.\ This model is based on the ideal gas law, i.e.\ independent of the gas species.}
\label{fig.onaxisdensity}
\end{center}
\end{figure}

A local density gradient imposes a varying refractive index $\eta$, resulting in a phase shift $\Delta \phi$ of rays passing through that particular region in comparison to rays which would have propagated through vacuum.\ Therefore, the undisturbed fringe spacing is locally shifted by $\Delta S$.\ The relationship between the atomic density $\rho$ and refractive index $\eta$ of a gas is given by the Lorentz--Lorenz equation \cite{liu}
\begin{equation}
\rho= \frac{\eta^2-1}{\eta^2+2}\frac{N_A}{A} \approx (\eta-1) \frac{2}{3}\frac{N_A}{A} \label{eq.LL} \text{,}
\end{equation}
where $N_A$ is Avogadro`s number and $A$ is the molar refractivity, e.\ g.\ $A_ {Ar}\approx4.20\times 10^{-6} \text{ m}^3/\text{mol}$.\ The approximation is valid for $\eta\approx 1$, which is fulfilled for a gas with densities $n<10^{19}\,/\SI{}{\cubic\centi\metre}$.\ Inverting equation \ref{eq.LL} yields the dependency of the refractive index on the density.\ The phase shift between interfering rays is given by $\Delta \phi = \phi(y) - \phi(y-d)$, where $\phi$ is the phase imposed by the neutral gas atoms and $y$ is the coordinate along the direction perpendicular to the interference fringes.\ Assuming the linear approximation (equation \ref{eq.LL}) a homogeneous density within a gas jet of diameter $l$ yields 
\begin{equation}
\phi \approx l (\eta-1)\frac{2\pi}{\lambda} \label{eq.dphi} = l \frac{3\pi}{\lambda} \frac{A}{N_A} \rho \text{.}
\end{equation}
\begin{figure}[h!]
\begin{center}
\includegraphics[width=6cm]{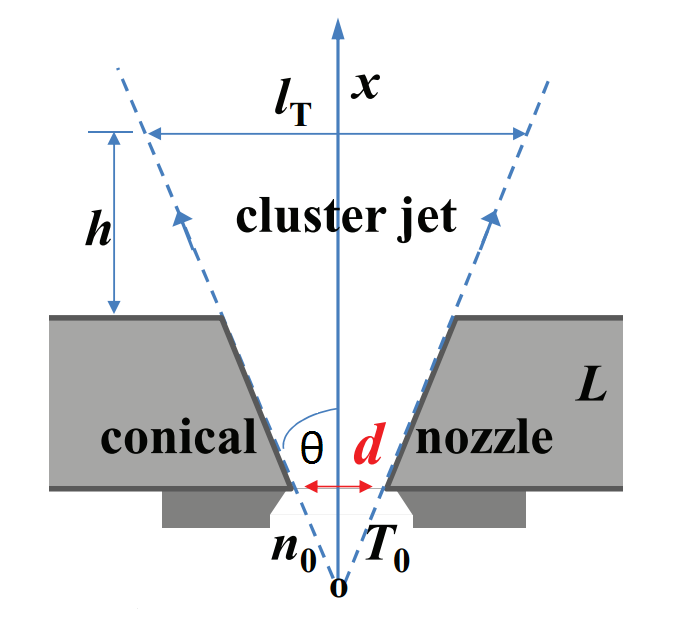}
\caption{Sketch of a conical nozzle \cite{chen}.\ The quantities for the solenoid valve are $d=\SI{500}{\micro\metre}$, $L=\SI{250}{\micro\metre}$ and $\theta=45^{\circ}$.}
\label{fig.nozzle}
\end{center}
\end{figure}
This expression yields an upper limit for the phase shift, since the gas jet has a circular throat and hence, produces a cylindrically symmetric gas flow, i.e.\ the path length of a ray propagating through the gas is at most $l$.\ As described above, the fringe spacing is affected by the phase shift in the FFW set-up; the fractional fringe distance shift is given by
\begin{equation}
\frac{\Delta S}{S}=\frac{\Delta \phi}{2\pi} \label{eq.dSS} \text{.}
\end{equation}
For comparison purposes, and because higher pressures than  \SI{15} .... \SI{20}{\bar} bar are possible, the gas density using the solenoid valve (Parker Miniature High--Speed Valve) will be estimated.\ The pressure of the gas before leaving the nozzle, the backing pressure $P_b$, as well as the nozzle design play a crucial role for the gas density distribution in the jet.\ The gas jet used in the present experiment allows for backing pressures up to 80\,bar and has a conical nozzle with a half opening angle $\theta=45^{\circ}$ (see figure\ \ref{fig.nozzle}).\ As described by Chen \cite{chen} the on--axis particle density at a height $x$ of a conical gas jet is approximately given by
\begin{equation*}
\frac{\rho}{\rho_0} \approx 0.15\left(\frac{0.74\,d}{x\,\tan\theta}\right)^2 \text{,}
\end{equation*}
where $x$ is the distance to the throat of the nozzle (see figure \ref{fig.nozzle}), $\rho_0$ is the atomic number density of the gas before leaving the nozzle and $d$ is the throat diameter of the valve.\ 
The gas jet used in this study has a throat diameter  $d=\SI{500}{\micro\metre}$ and $L=\SI{250}{\micro\metre}$.\ Using the ideal gas law, which is a good approximation for the monoatomic gas argon at room temperature, $\rho_0$ is determined by the backing pressure $P_b$, which refers to the pressure in the chamber before the nozzle as well as the temperature of the nozzle $T_0$ via: $\rho_0=P_b/(k_B T_0)$, where $k_B$ is the Boltzmann constant.\ Figure \ref{fig.onaxisdensity} shows the on-axis density $\rho$ with respect to $h$ (distance from the end of the nozzle) for different backing pressures from 10 to 40 bar for the dimensions of the Parker solenoid valve.\ 

A practical example for gas jets to produce a plasma for LWFA: in order to excite the plasma wake resonantly with a laser pulse  of 15\,fs duration an electron density in the order of \SI{1.75e18}{\per\cubic\centi\metre} is necessary \cite{nick}.\ The required density (assuming argon being ionized eight fold) at $h=2.5\,$mm (distance from the nozzle) is achieved with backing pressures of \SI{30}{\bar}.\ According to equation \ref{eq.dphi} the phase shift of a gas jet with an average density ranging from \SIrange{1e18}{1e19}{\per\cubic\centi\metre} lies between \SI{0.1}{\radian} and \SI{1}{\radian}.\ A phase shift of this order is expected to result in a clearly visible change of the interference pattern described by equation \ref{eq.dSS}.\ 

\subsection{Experimental Set-Up for Gas Density Measurements} 
The Wollaston interferometer to measure the gas density is set up on a single breadbord (\SI{75}{\centi\metre}$\times$\SI{125}{\centi\metre}), such that the optical set-up and vacuum chamber with the gas jet form a compact and transportable unit.\

\begin{figure}[h!]
\begin{center}
\includegraphics[width=\textwidth]{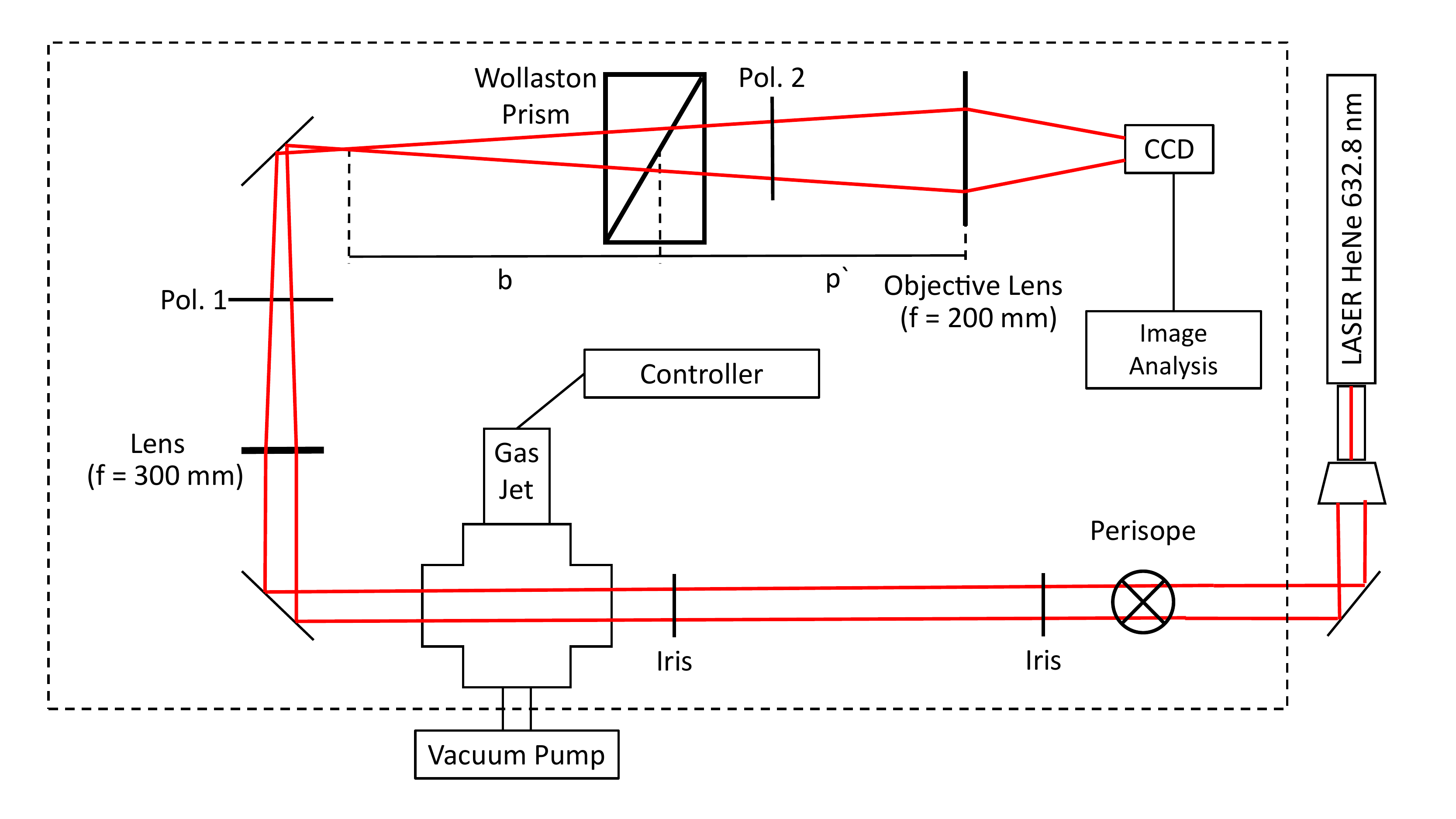}
\caption{Sketch of the interferometric set-up using a Wollaston prism.\ The dashed line represents the breadboard with the shielding box.\ The Wollaston prism is mounted on a linear stage, such that the parameter $b$ and therefore the fringe spacing $S$ can be changed (see section \ref{sec.wprism}).\ A Nikon imaging lens is attached to the CCD camera.}
\label{fig.setup1}
\end{center}
\end{figure}

\subsubsection{Wollaston Interferometer} \label{sec.expwollaston}
\begin{figure}[t!]
\begin{center}
\includegraphics[width=\textwidth]{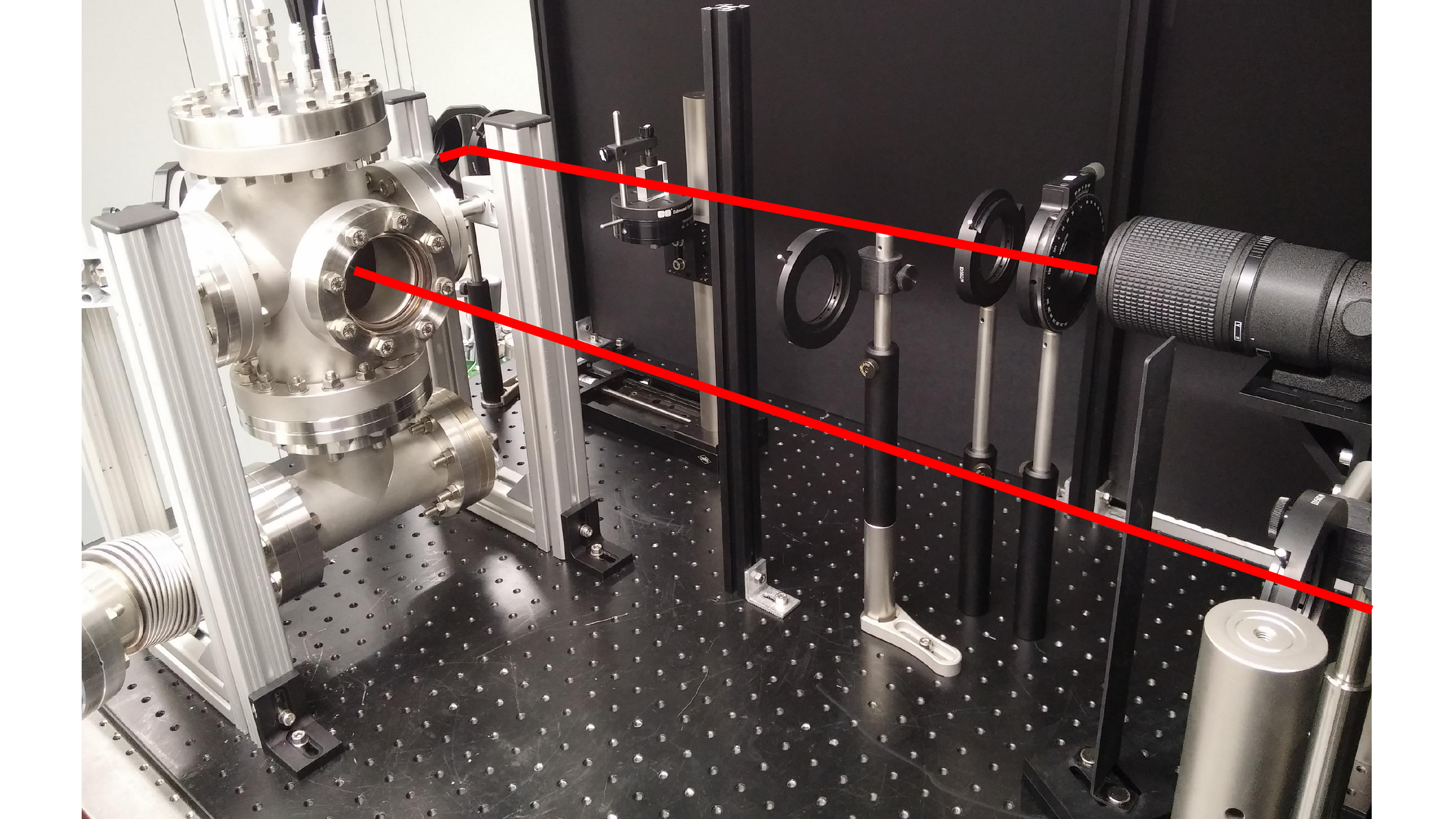}
\caption{Experimental set-up for Wollaston interferometry.\ The expanded beam passes through the gas jet, the first polarizer, the lens, the Wollaston prism, and the second polarizer. The lens images the gas jet onto the CCD by a 200 mm imaging lens (AF Micro--Nikkor 200 mm f/4D IF--ED, Nikon).}
\label{fig.setup2}
\end{center}
\end{figure}

\begin{figure}[b!]
\begin{subfigure}[b]{0.45\textwidth}
\includegraphics[width=\textwidth]{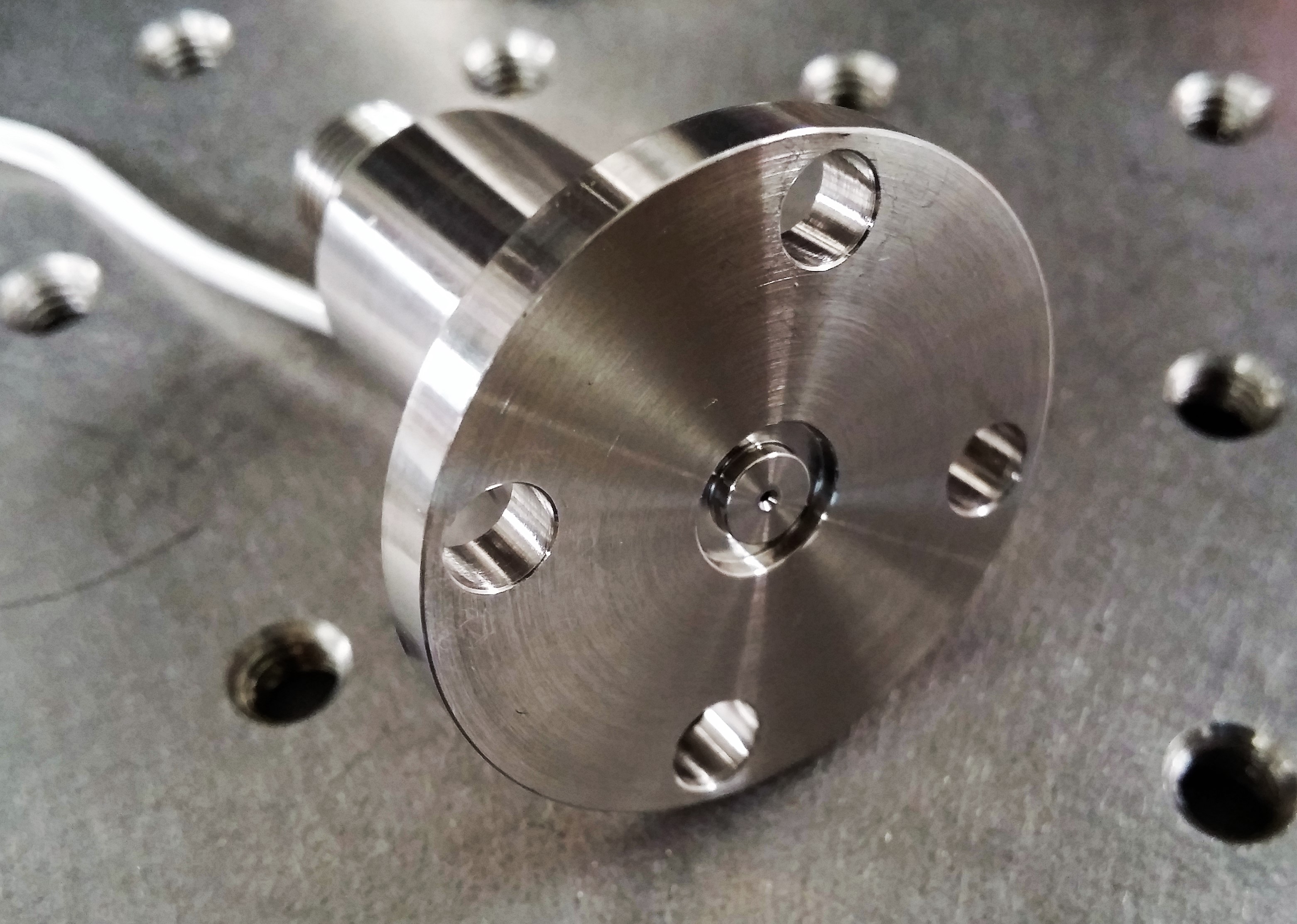}
\caption{Miniature high speed high vacuum dispense valve with conical outlet (Parker 009--0442--900).}
\label{fig.valve}
\end{subfigure}
\hfill
\begin{subfigure}[b]{0.45\textwidth}
\includegraphics[width=\textwidth]{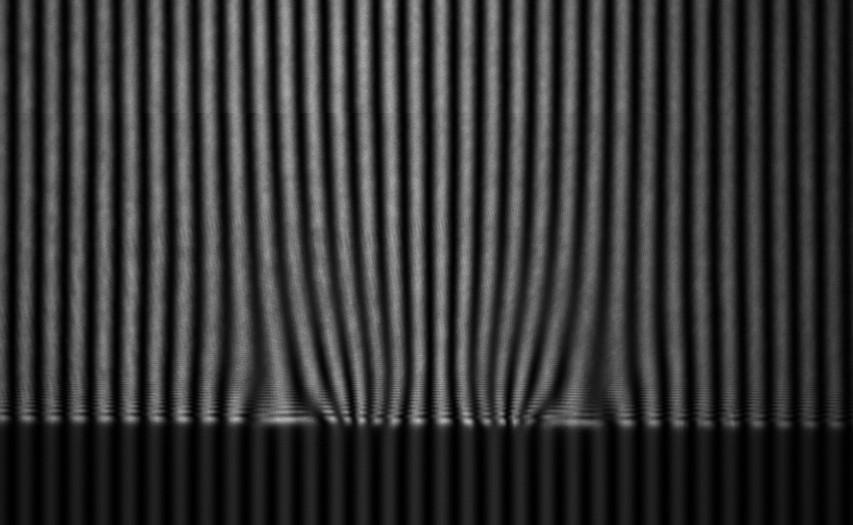}
\caption{Typical interference fringes of Parker solenoid valve at 35 bar backing pressure with argon.\ Here, the gas flow is directed upwards. The two non-straight areas correspond to positive and negative phase differences $\Delta \phi$ (see equation \ref{eq.dSS}).
}
\label{fig.rawfringe}
\end{subfigure}
\caption{Parker solenoid valve and interference fringes.}
\end{figure}

The interferometer is depicted schematically in figure \ref{fig.setup1} and figure \ref{fig.setup2}.\ A linearly polarized continuous He:Ne laser with a wavelength $\lambda_I=\SI{632.8}{\nm}$ and an output power of 21 mW is used as a coherent light source for the interferometer.\ To prevent upstreaming air due to the heat of the laser from causing unwanted phase shifts the laser is placed outside of the black cardboard box that contains the interferometer.\ The box reduces noise as it minimises the airflow in the whole experiment, as well as external photons hitting the CCD sensor.\ The initial laser diameter is 0.7\,mm (1/$e^2$ width).\ The telescope (20$\times$) attached to the laser provides a beam with a diameter of 14\,mm, which is suitable to backlight a gas jet with a diameter of a few millimetres.\ The noise of the acquired images is expected to be lowest in the centre of the beam due to the higher intensity of the laser in that region.\ The He:Ne laser enters and leaves the vacuum chamber through two windows, 
the gas flows in perpendicular direction, downwards towards the vacuum pump.
 The Parker solenoid valve (figure \ref{fig.valve}) is operated at 0.5 to 3\,Hz and with opening times $T\leq\SI{12}{\milli\s}$ to reduce the gas load on the pump.\ The Wollaston prism is placed between two crossed polarizers, such that the interference fringes are parallel to the jet.\ The interference pattern is captured with a CCD camera and a f = 200 mm camera
  lens.\ The minimal exposure time of the camera is \SI{18}{\micro\s}.\ The exposure time is set to \SI{30}{\micro\s} in order to obtain an image with good contrast without reaching saturation of the sensor.\ 


Figure \ref{fig.rawfringe} shows a typical interference pattern of the undisturbed gas jet.\ The change of the fringes due to the gas distribution is clearly observable when the Parker solenoid valve is operated at 35 bar backing pressure.\ 

\subsection{Experimental Set-Up for Non--Rotational Measurements}
\begin{figure}[h!]
\begin{center}
\includegraphics[width=\textwidth]{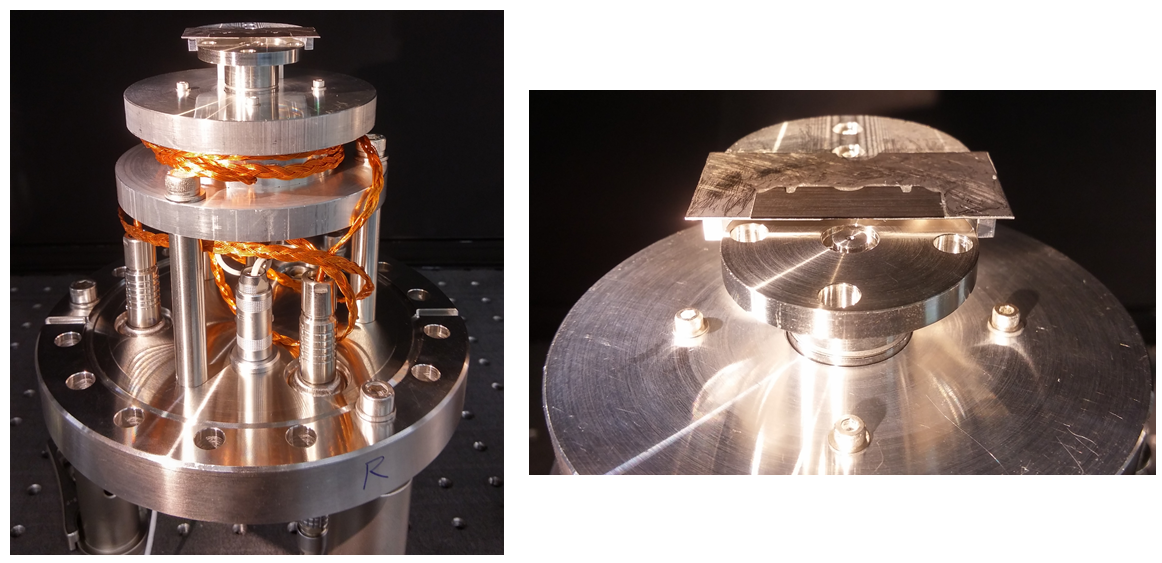}
\caption{Tomography set-up for shock front studies.\ A razor blade is positioned with a linear and a rotational stage.}
\label{fig.tommodule}
\end{center}
\end{figure}

Figure \ref{fig.tommodule} depicts the experimental set-up to obtain tomographic projection data of the solenoid gas jet with a supersonic shock front generated by inserting a razor blade into the gas flow.\ In order to to keep the gas jet and the imaging system fixed but to position the blade around the gas jet with two degrees of freedom.\ This is achieved with a rotational piezo stage 
and a linear piezo positioner, 
which is attached to the rotational stage.\ The rotational stage allows to measure phase projections along different directions with respect to the orientation of the shock front.\ The resulting density gradient is shown in Figure \ref{fig.ramp_ana}. The radial positioning can then be used to vary the position of the blade with respect to the center of the gas jet.\ This allows us to change the properties of the shock front.\ Furthermore, the radial degree of freedom can be used to correct the eccentricity between the centre of the gas jet and the axis of the rotational stage due to mechanical imperfections.\ The vertical distance between the razor blade and the nozzle of the gas jet is fixed in this set-up to \SI{1.6}{mm}.\   

\section{Data Analysis for Interferometry and Real-Time Tomography} \label{sec:analysis}
The Wollaston interferometry set-up introduced in the previous section can be used to measure phase projections along the propagation direction of the interferometry laser.\ An optically transmitting object that has a refractive index close to 1 (e.\ g.\ a gas jet) can be characterised.\ If the studied object can be assumed to be rotationally symmetric around a central axis, an Abel inversion yields the 3D density distribution from a single phase projection, i.e. a 2D projection obtained in one single observation direction, which is perpendicular to the axis of symmetry.\ In order to obtain the 3D density distribution of an object without such symmetry (e.\ g.\ a gas jet with a shock front) several projections along different angles are needed.\ The 3D distribution can then be reconstructed with tomographic algorithms, for instance the Maximum Likelihood Expectation Maximisation (ML--EM) c.f.\
section \ref{sec.mlem}.\
The images showing the interference fringes obtained with the Wollaston interferometer (section \ref{sec.expwollaston}) need to be evaluated numerically in order to obtain quantitative information about the phase projections, and hence to be able to reconstruct the asymmetric, spatial density distribution.

\subsection{Numerical Tools} \label{sec.fringeana}
The undisturbed fringes are mathematically described by a harmonic oscillation with a fixed frequency multiplied by the Gaussian amplitude of the laser beam profile.\ The wave period of the oscillation is given by the undisturbed fringe width $S$ (equation \ref{eq.fringe_width}), i.e. the spacing between the fringes.\ A gas jet will locally change the fringe width according to the phase introduced to the different light rays in the laser beam by the gas flow.\ The following phase extraction problem arises \cite{wrap}: given noisy discrete values of a function of the form
\begin{equation*}
g(y)= b(y) \cdot e^{\mathrm{i}[2\pi f y + \alpha(y)]} \text{,}
\end{equation*}
where $\alpha(y)$ represents the phase shift induced by the gas jet.\
The term $b(y)$ takes global intensity changes due to the Gaussian profile of the interferometry laser into account, $y$ is the coordinate perpendicular to the gas flow and interferometry laser and $f$ is the oscillation frequency of the undisturbed fringes.\ This problem can be solved by fitting the measured data to an ansatz for $\alpha(y)$.\ However, this is not appropriate, since one has to make (possibly incorrect) assumptions about the form of $\alpha(y)$.\ This is particularly true, when a shock front has to be characterised by $\alpha(y)$.\ A more elegant way to directly unwrap the phase from the noisy data makes use of a Fourier transformation
\begin{equation*}
\mathcal{F}(g)(\omega) = \int_{-\infty}^{+\infty} e^{-\mathrm{i} 2\pi \omega y} \cdot b(y) \cdot e^{\mathrm{i}[2\pi f y + \alpha(y)]}\ \mathrm{d}y \text{.}
\end{equation*}
Since the acquired signal is real, the full information is contained in the positive frequency domain where the spectrum has two significant peaks.\ One at the fringe frequency $f$ and another around zero 
due to the slowly varying Gaussian intensity profile of the beam.\ A Gaussian window is applied to the spectrum to cut out the low frequency peak.\ In order to eliminate the $2\pi f y$ phase factor we apply a rotation of the Fourier--transformed signal by
\begin{equation*}
R_f \mathcal{F}(g)(\omega)=\mathcal{F}(g)(\omega+f) = \int_{-\infty}^{+\infty} e^{-\mathrm{i} 2\pi \omega y} \cdot b^\prime(y) \cdot e^{\mathrm{i} \alpha(y)}\ \mathrm{d}y\text{,}
\end{equation*}
with the result that the rotation by $R_f$ shifts the peak at frequency $f$ to zero.\

In order to obtain the sought phase shift $\alpha$, we apply the inverse Fourier--transform
\begin{equation*} 
\mathcal{F}^{-1}(R_f \mathcal{F}(g))(y)= b^\prime(y) \cdot e^{\mathrm{i} \alpha(y)} =: A (y)\text{.}
\end{equation*} 
As a result, $A$ contains the information about the phase shift $\alpha$ as well as remaining effects of $b$ denoted by $b^\prime$.\ These are minor effects due to imperfection at mirrors, lenses and polarizers or inhomogeneities of the two windows the laser passes through.\ These effects are eliminated by 
the point wise product $A\cdot A_{ref}^{-1}$, where $A_{\text{ref}}$ is the signal obtained by the same transformation applied to the reference data, i.e.\ without gas jet.\ The phase of the remaining term corresponds to the phase caused by the gas jet alone.\ 

The term $e^{\mathrm{i}2\pi f y}$ would arise in $A_{\text{ref}}$ as well and is therefore canceled in the last step, even if the rotation $R_f$ is omitted.\
As explained in section \ref{sec.wprism}, the Wollaston interferometer measures the phase difference between rays separated by a distance $d$ in the object plane.\ Therefore, phase shifts obtained at two points in the image plane that are separated by a distance corresponding to $d$ have to be added up.\
We obtain the additional phase $\phi$ by
$
\phi(y)= \frac{2 \pi}{\lambda}\int (\eta(y,z)-1)\ \mathrm{d}z \text{.}\
$
This is the phase which a ray has picked up when passing through the gas jet, in comparison to a ray that has passed through vacuum only.\ The integration is taken along the ray's path at position $y$, $\lambda$ is the wavelength of the laser and $\eta(y,z)$ refers to the distribution of the refractive index in the gas jet.\ As long as cylindrical symmetry of the gas distribution is assumed, obtaining the radial density distribution $\rho (r)$ from $\phi (y)$ is possible.\ The corresponding integral relation is called inverse Abel transformation or Abel inversion. The phase shift $\phi$, extracted with the method explained before, is proportional to a projection of the gas jet along the propagation direction of the laser.\ Here, cylindrical symmetry of the gas distribution is assumed.\ This assumption enables to reconstruct the radial density distribution from $\phi$.\ This is achieved by the inverse Abel transform \cite{abel2,abelwiki}, which reads

\begin{equation*}
\phi(y)=\frac{\Delta s}{\lambda}2 \pi= \frac{2\pi}{\lambda} 2 \int_0^\infty \left[\eta(y,z)-1\right]\,\mathrm{d}z =\frac{4\pi}{\lambda} \int_y^\infty \left[\eta(r)-1\right]\frac{r}{\sqrt{r^2-y^2}}\mathrm{d}r\text{,}
\end{equation*}

\begin{equation*}
F(y):= \phi(y) \cdot\frac{\lambda}{2 \pi} = 2 \int_y^\infty \left[\eta(r)-1\right]\frac{r}{\sqrt{r^2-y^2}}\mathrm{d}r\text{,}
\end{equation*}

\begin{equation}
\label{eq.abel}
\overset{\text{inv. Abel}}{\implies} \eta(r)-1 = -\frac{1}{\pi} \int_r^\infty \frac{\mathrm{d}F(y)}{\mathrm{d}y}\frac{\mathrm{d}y}{\sqrt{y^2-r^2}} \text{.}
\end{equation}
The derivative and the integral occurring in equation \ref{eq.abel} are calculated following \cite{abel}.\ The key points are summarised here.\ It is not suitable to approximate the derivative of $F$ by the discrete differential quotient due to noise, since this would amplify noise drastically.\ A more sophisticated method based on Gaussian filters is needed.\ Consider the following Fourier identity for $f$, $g$ functions with compact support.
\begin{align}
\label{eq.f1}
\mathcal \mathcal{F}(f)\cdot\mathcal{F}\left(\frac{\mathrm{d}g}{dx}\right)   &= \mathcal \int_{-\infty}^{+\infty} \mathrm{d}x\ e^{-\mathrm{i}2\pi x\omega} f(x) \cdot \int_{-\infty}^{+\infty} \mathrm{d}x\ e^{-\mathrm{i}2\pi x\omega} \frac{\mathrm{d}g(x)}{dx}  \\
&=\mathcal \int_{-\infty}^{+\infty} \mathrm{d}x\ e^{-\mathrm{i}2\pi x\omega} f(x) \cdot \int_{-\infty}^{+\infty} \mathrm{d}x\ e^{-\mathrm{i}2\pi x\omega}\ (\mathrm{i}2 \pi \omega)\ g(x) \nonumber  \\
&=\mathcal \int_{-\infty}^{+\infty} \mathrm{d}x\ e^{-\mathrm{i}2\pi x\omega}\ (\mathrm{i}2 \pi \omega)\ f(x) \cdot \int_{-\infty}^{+\infty} \mathrm{d}x\ e^{-\mathrm{i}2\pi x\omega} g(x)   \nonumber \\
&=\mathcal \int_{-\infty}^{+\infty} \mathrm{d}x\ e^{-\mathrm{i}2\pi x\omega} \frac{\mathrm{d}f(x)}{dx} \cdot \int_{-\infty}^{+\infty} \mathrm{d}x\ e^{-\mathrm{i}2\pi x\omega} g(x)    \nonumber \\
&=\label{eq.f5} \mathcal \mathcal{F}\left(\frac{\mathrm{d}f}{dx}\right)\cdot\mathcal{F}(g)    \nonumber
\end{align}
When $g$ is set to a Gaussian distribution, rearranging this identity yields the smoothed derivative of a noisy signal $f$ by using only the derivative of $g$ instead of $f$ (equation \ref{eq.f1}).\ The derivative of the Gaussian function $g$ can be easily evaluated from its analytical expression.\ Another issue is the singularity inside the integral of equation \ref{eq.abel}, when the integral is approximated on a discrete domain.\ This is solved by setting the first value of the integral ($y=r$) to the second value ($y=r+\Delta y$).\ For the limit of many points (i.e.\ infinitesimal grid spacing $\Delta y$) the numerical value of the integral will converge to the analytical value \cite{abel}.\ 

\subsection{Tomographic Reconstruction}

The goal of tomographic reconstruction algorithms is to estimate a 3D density distribution based on its measured projections along directions with different angles.\ If rotational symmetry can be assumed an Abel inversion provides an analytic method to reconstruct the 3D density distribution from a single projection.\ Problems where rotational symmetry cannot be assumed demand for another reconstruction method.\ These kind of problems arise frequently in medical physics when a density image of tissue is desired, but only projections from x--ray scans or PET (positron emission tomography) data is available.\ An efficient algorithm is the so called Back Projection.\ As computation power increased drastically, more sophisticated algorithms were developed.\ One of these is the Maximum Likelihood -- Expectation Maximization (ML--EM) algorithm \cite{tomo}, which is an iterative method with good convergence properties compared to the Back Projection method but, on the other hand, has a higher computational cost.\ 

The idea of a simple Back Projection is to redistribute (back--project) the measured projections homogeneously along the projection lines.\ This procedure can be implemented efficiently but it can only provide rough information about the distribution.\ The reconstructed images become blurred, albeit infinite projections are available \cite{tomo}.

\subsubsection{Maximum Likelihood -- Expectation Maximisation (ML--EM)}\label{sec.mlem}

\begin{figure}[h!]
  \begin{subfigure}[b]{0.45\textwidth}
    \includegraphics[width=\textwidth]{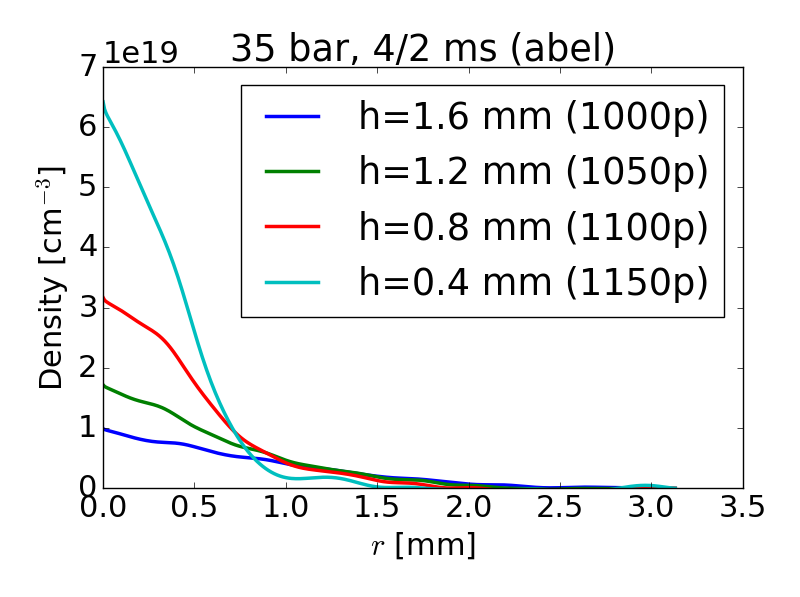}
    \caption{Abel inversion}
  \end{subfigure}
  \hfill
  \begin{subfigure}[b]{0.45\textwidth}
    \includegraphics[width=\textwidth]{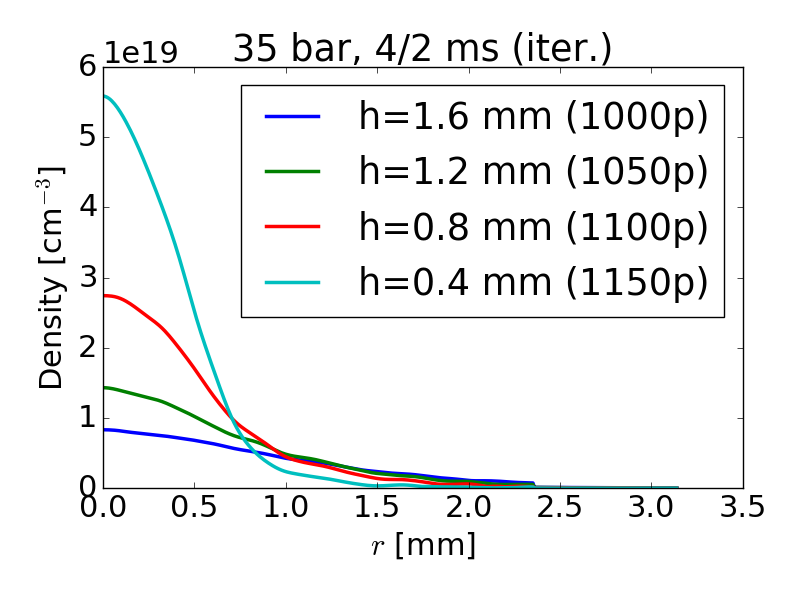}
    \caption{ML--EM}
  \end{subfigure}
  \caption{Test of ML--EM on the undisturbed (rotationally symmetric) gas distribution.\ The distance from the nozzle is given in mm; the corresponding pixel row is placed in brackets.\ In general, good agreement is found between the Abel inversion (a) and ML--EM after 15 iterations (b).\ The problem due to the singularity of the Abel inversion at $r=0$ does not arise with ML--EM.}\label{fig.radial_test}
\end{figure}

\begin{figure}[h!]
    \centering
    \includegraphics[width=\textwidth]{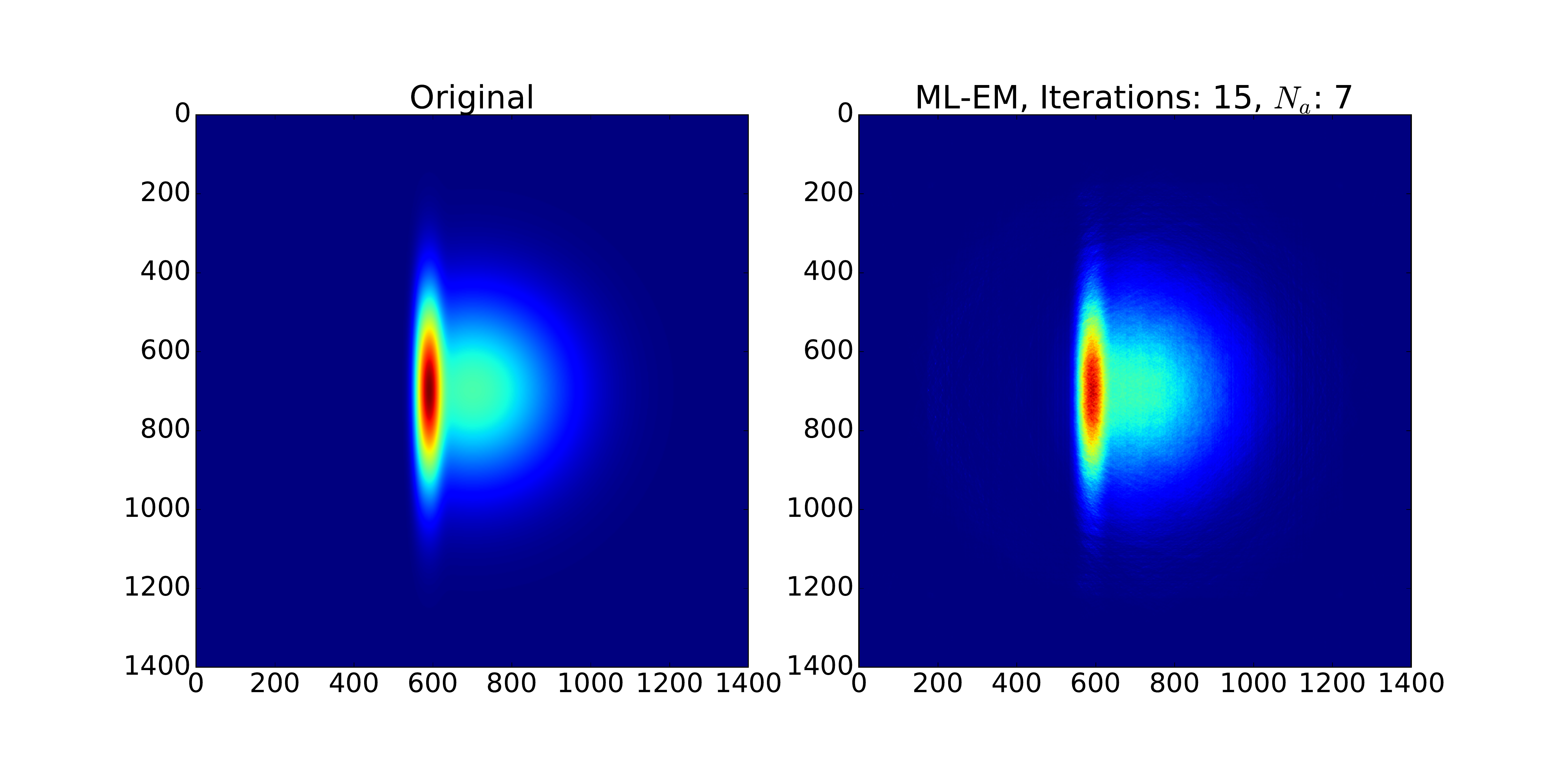}
    \caption{Reconstruction by ML--EM from 7 projections with artificial gaussian noise, $\sigma=0.01$.}
    \label{fig.mlem_example} 
\end{figure}

The basic principle of this algorithm is to advance an initial guess of the distribution iteratively by comparing the forward--projected data of the current guess with the measured data from all angles.\ The $n$--th estimate of the $i$--th voxel is named $x_i^n$.\ Often, a homogeneous distribution is chosen as the initial guess $x_i^0$ .\ The ML--EM step that advances the guess is computed by the following equation \cite{tomo}
\begin{equation*}
x_i^{n+1}=x_i^n \frac{1}{\sum_j A_{ij}} \sum_j A_{ij} R_j^n \text{,}
\end{equation*}
where $R_j^n$ is the ratio of the value of measurement pixel $y_j$ and the forward projected data of the $n$--th estimate
\begin{equation*}
R_j^n=\frac{y_j}{\sum_k A_{kj} x_k^n} \text{.}
\end{equation*}
The matrix $A$ is the system matrix, which accounts for how much each voxel contributes to each measurement.\ By this procedure, the advanced guess will approach a distribution that is compatible with all measurements.\ More precisely, the computed guess converges to a distribution that has a high probability (maximum likelihood) to be the original distribution, given the measured data.\ 
First, the ML--EM algorithm is applied to a single phase projection from the undisturbed gas jet.\ The reconstructed radial density distribution agrees well with the result from Abel inversion (figure \ref{fig.radial_test}).\ The Abel inversion shows non--smooth behaviour around $r=0$ due to the singularity.\ This problem does not arise with the ML--EM algorithm.\ 
To validate the performance of the algorithm on non-rotational symmetric distributions, tests with known distributions are carried out.\ A 2D Gaussian multiplied by a step function is chosen, as the shock--wave by the razor blade is expected to have a similar shape.\ Figure \ref{fig.mlem_example} shows the generic distribution, as well as the reconstructed image from $N_a=7$ projections after 15 iterations.\ Random noise distributed according to a Gaussian distribution with a standard deviation of $\sigma=0.01$ is added to the projection data before running the ML--EM reconstruction and qualitative reconstruction of the original distribution without rotational symmetry is achieved.\

\subsubsection{Convergence and Error Studies of ML--EM}\label{sec.mlem_test}

\begin{figure}[h!]
    \centering
    \includegraphics[width=.5\textwidth]{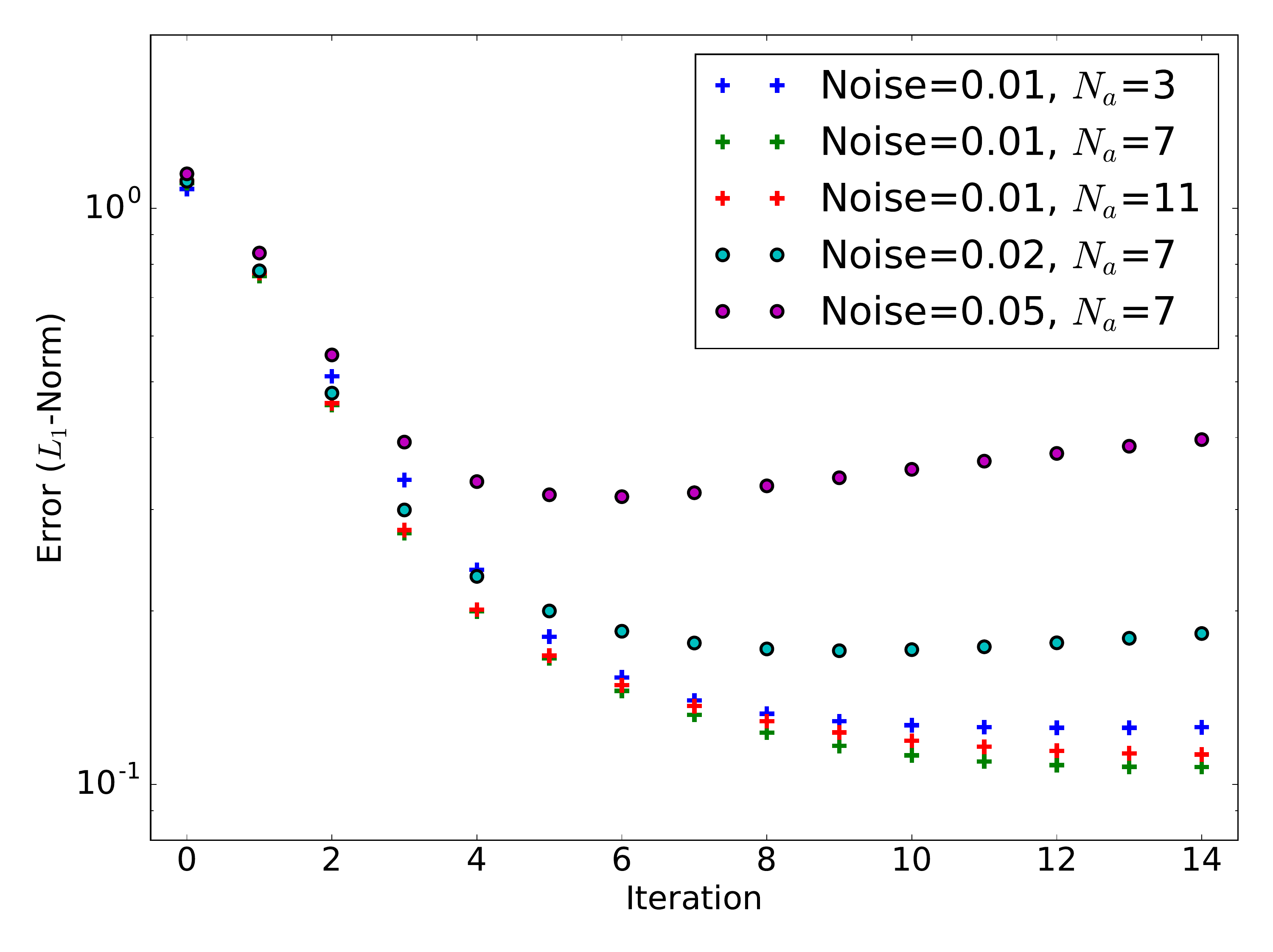}
    \caption{Convergence studies of ML--EM for various degrees of Gaussian noise $\sigma$ and number of projection angles $N_a$.\ Convergence is observed after 7--10 iterations.}
    \label{fig.mlem_error} 
\end{figure}

The convergence properties of the ML--EM algorithm are summarised in figure \ref{fig.mlem_error}, which shows the $L_1$  norm for the first 15 iterations for different numbers of measurements (number of projection angles $N_a$).\ Gaussian noise (standard deviation $\sigma=$ 0.01--0.05) is added to the measurement.\ The reconstructed distribution matches the original distribution best if noise is lowest and, more interestingly, the error is not strongly correlated to the number of projections.\ The downside of many projection angles is that more noise is picked up.\ It turns out that $N_a$ around 7 achieves best convergence properties for the distribution and a noise level of $\sigma=0.01$. 

\subsection{Real-Time Computation of the Density Reconstruction}
Real time tomographic reconstruction requires large computational resources.\ Since the ML--EM algorithm can be parallelised, real time reconstruction can be achieved using either multi-core CPUs or graphical processing units (GPUs).\ 

\begin{figure}[h!]
	\centering
	\includegraphics[width=.5\textwidth]{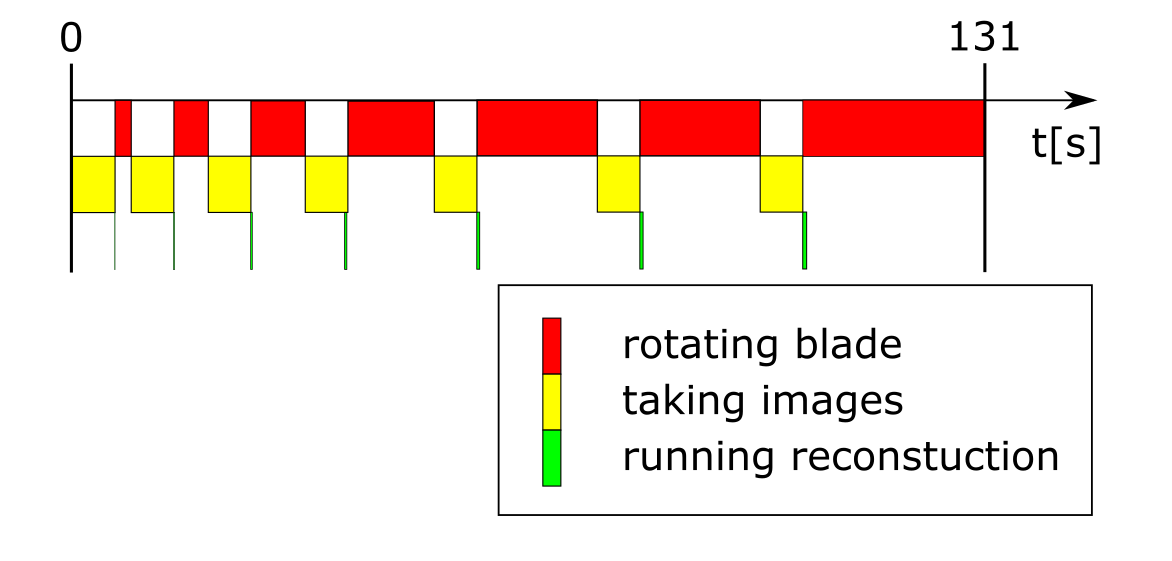}
	\caption{Time line of density reconstruction with $N_a = 7$ projection measurements. 
	The angles are chosen such that more projections are acquired in the direction parallel to the shockfront, maximising the information about the shock. For each projection 20 images were acquired, 10 of which were recorded with the gas jet on and the other 10 with the gas jet off, as a reference.}
	\label{fig.exp_timing} 
\end{figure}

The measurement can be divided into three different parts as sketched in figure\ \ref{fig.exp_timing}. In the first step, the actual image is taken. The blade is held at a fixed position, and the gas jet opens.\ In this experiment, the gas jet's open--close frequency is limited by the vacuum pump to about \SI{2}{Hz}.\ Since we average over 10 shots of the gas jet, to improve the statistics, the time needed for each step in angle is fixed to approximately \SI{5}{\second}.\ Between two projection measurements at different angles the blade has to be moved (second step). The maximal angular velocity is \SI{1}{deg\per \second}, hence for a complete measurement (\SI{90}{deg}) the total driving time is about \SI{90}{\second}.\
The last step of the measurement is the tomographic reconstruction of the density distribution. Since driving the blade does not require any computational resources, this process can be done in parallel. As soon as the stage is moving, the computation of the density distribution can be started. After approximately \SI{10}{\second} the first estimate of the 2D density distribution is available and improves consecutively with the number projections from different angles. The time required to reconstruct the density for $N_a = 7$ projections is below \SI{2}{\second}.


In order to create the parallel implementation of the algorithm, the TomoPy package for Python \cite{tomopy} together with the Astra toolbox \cite{astra} was used.\ TomoPy is a Python based framework for tomographic image reconstruction and data processing tasks developed at the Advanced Photon Source of Argonne National Laboratory.\ TomoPy includes many functions to perform pre-processing and image reconstruction using different algorithms \cite{tomopy}.\ The Astra toolbox is an open-source project developed at the University of Antwerp. The toolbox provides tomographic image reconstruction of 2D and 3D data sets. The toolbox uses CUDA to offload the reconstruction algorithms to the NVIDIA GPUs, but most of the 2D algorithms can also be executed on the CPU \cite{astra}.

Using the TomoPy toolbox the algorithm described in section \ref{sec.mlem} is implemented in Python to perform the image reconstruction. Additionally, the Astra toolbox is used in order to target the GPU platform. The GPU platform is mostly targeted in order to evaluate the potential to perform real-time 3D image reconstruction as for the 2D reconstruction the power of the CPU is mostly sufficient.

\begin{figure}[h!]
\centering
\includegraphics[width=.6\textwidth]{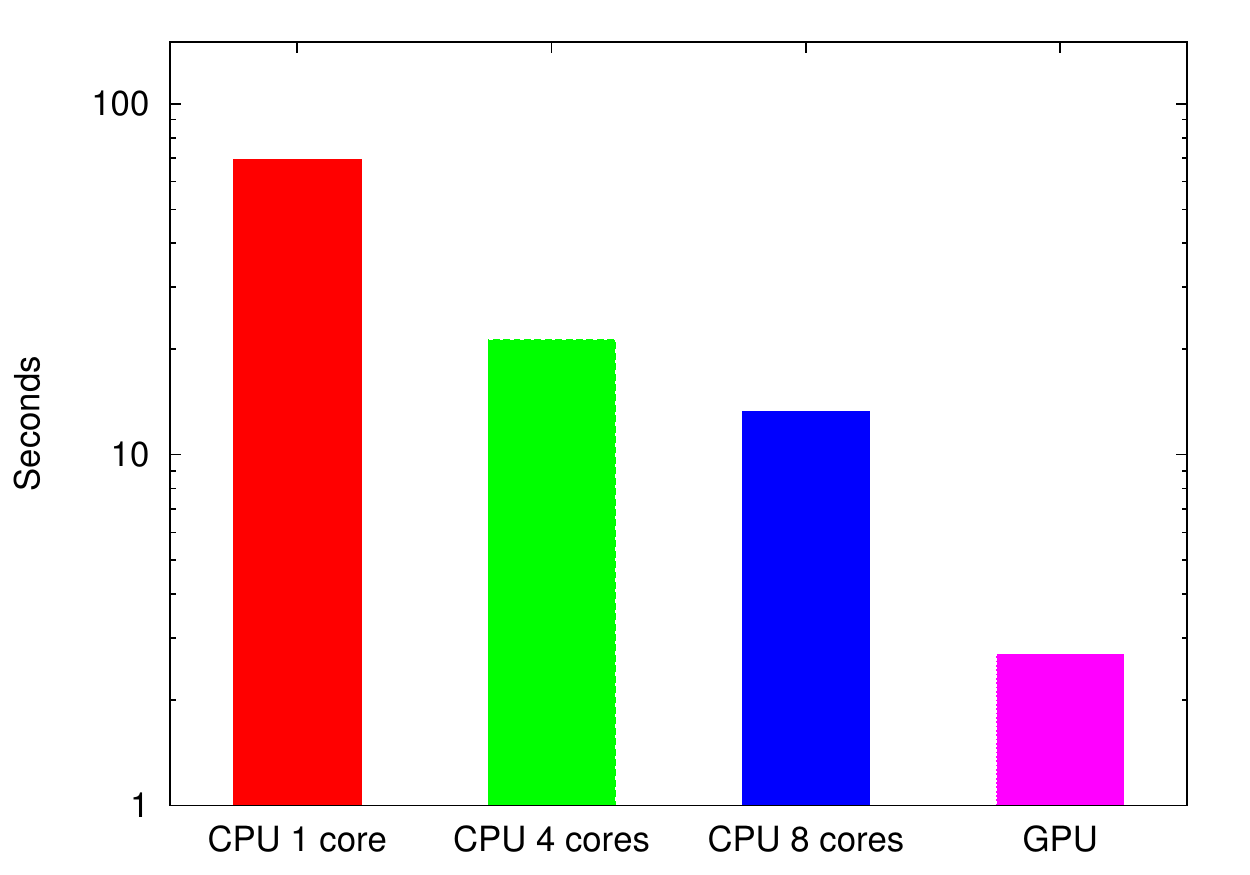}
\caption{Reconstruction times and speed-up using CPU, multi core CPU and GPU}
\label{fig:results}
\end{figure}


Reconstruction times of the algorithm are shown in the figure \ref{fig:results}.\ The figure shows reconstruction times using 1, 4, and 8 CPU cores as well as reconstruction offloaded to the GPU.\ The hardware used in the reconstruction was 2x Intel Xeon E5-2609 v2 CPUs and 1x NVIDIA Tesla K40c.\ The time represented in the figure shows only the reconstruction time without input and output operations, which are constant for all the implementations.\ The benchmark was done for a 3D reconstruction using 100 slices. 

With these technologies, the bottle neck of the experiment is clearly not the computational part anymore. Even when, for example a stage like the U-651 from Physik Instrumente (PI) with a maximal speed of \SI{540}{deg/s} would have been used, the time needed to drive the blade could be pushed below one second, still on the same order as the computation.


The achieved speed-ups on multi-core CPUs and GPU will allow to perform reconstruction in real time.\

\section{Examples Of Gas Jet Density Measurements}\label{sec:ExampleMeasurements}

In this section, two different argon gas jets (piezo-driven or solenoid-based valves) are studied under various conditions.\ A shock front arising from a razor blade inserted into the gas flow of the solenoid jet is evaluated via tomography.\ 

\subsection{Piezo gas jet for Free Electron Lasers beam instrumentation } 
Gas-based monitors are attractive for beam instrumentation, as they can be used to build quasi-noninvasive diagnostics in free electron lasers. The low density of the active medium allows the electrons and photons, respectively, to pass the monitor with only minimal impact on the beam.
The gas atoms are ionized by the beam, and the photoelectrons and / or ions can be characterized by appropriate detectors.
Piezo gas jets can deliver these atoms directly into the beam line vacuum. When combined with a turbo-molecular vacuum pump and used with microsecond opening times \cite{piezo7us}, these systems can be operated without the need for beam windows.

\begin{figure}[h!]
    \centering
    \includegraphics[width=0.45\textwidth]{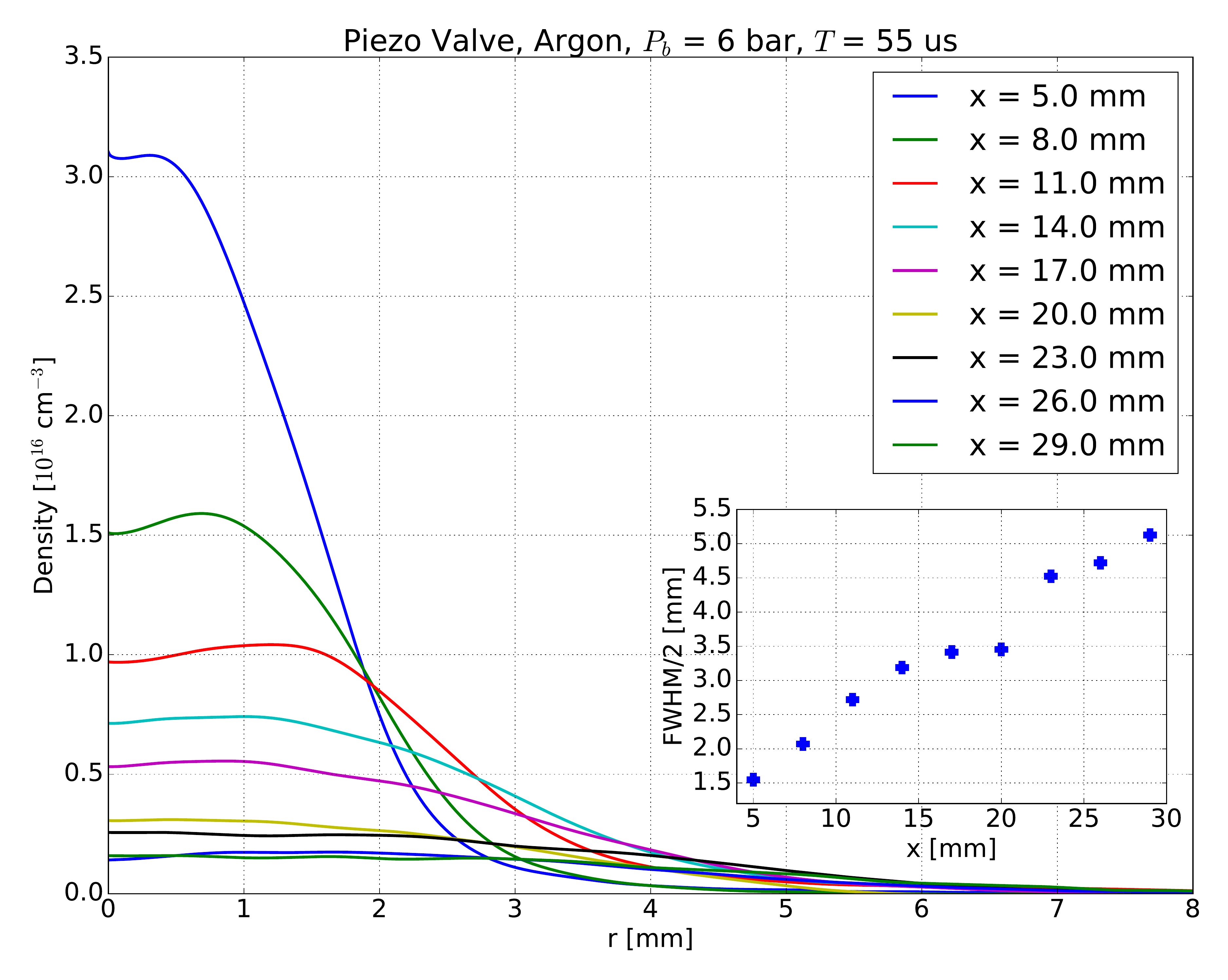}
    \caption{Density measurements of the piezo valve used for SwissFEL instrumentation.\ The reconstruction is computed via an Abel inversion. The subplot indicates the
    FWHM of the gas density distribution as function of the distance from the throat.}
    \label{fig.piezo_radial} 
\end{figure}

\begin{figure}[h!]
    \centering
    \includegraphics[width=0.45\textwidth]{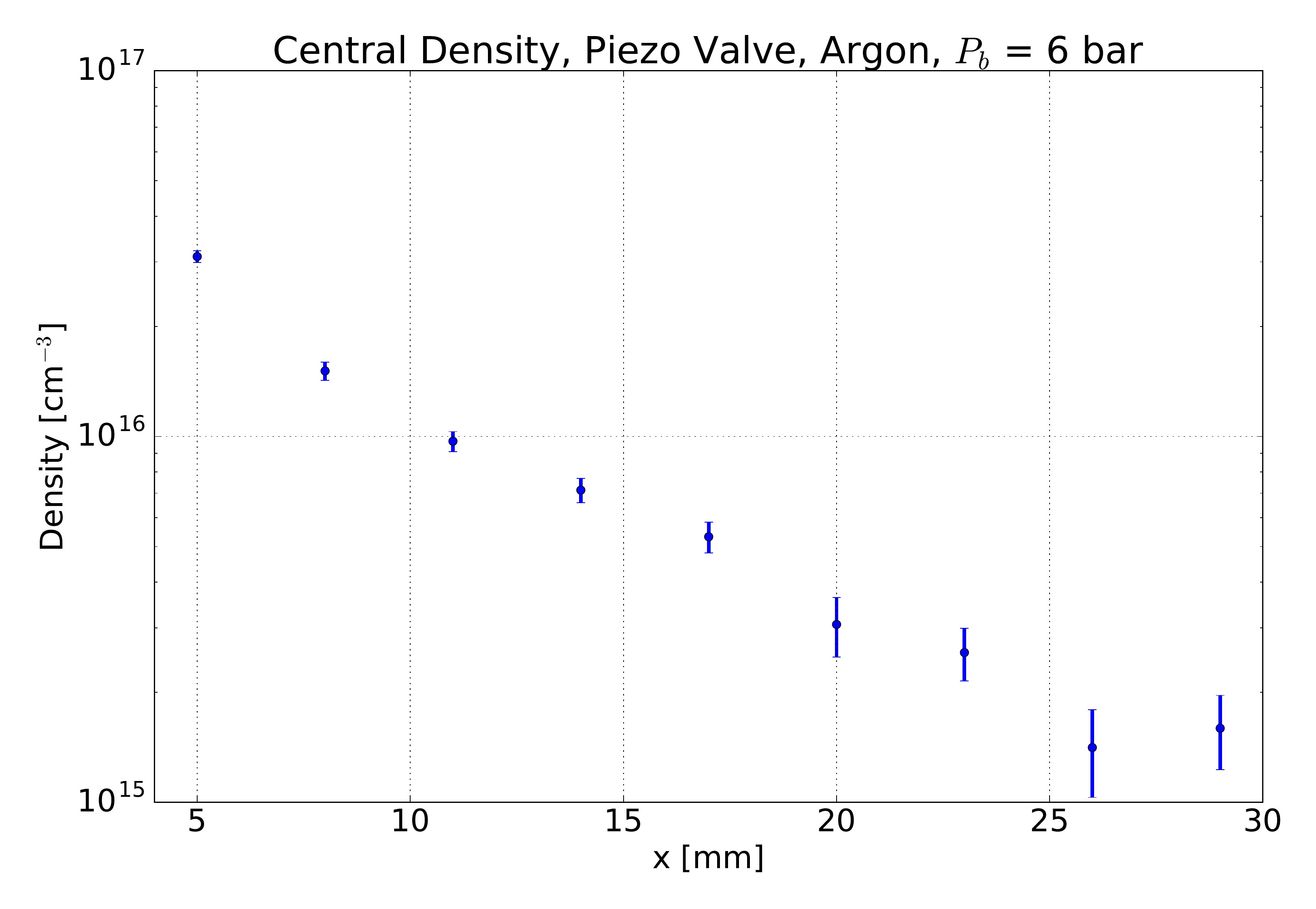}
    \caption{The central density in the gas jet reduces by a factor of 20 after expanding \SI{25}{mm}.}
    \label{fig.piezo_cendens} 
\end{figure}

SwissFEL is equipped with a Terahertz streak camera for measurements of the X-ray pulse length and arrival time \cite{THzSCOverview}. In this device, the X-ray pulses ionise xenon atoms, and the photoelectrons are streaked by an externally applied Terahertz field \cite{THzSCIntro}.
For the X-ray pulse length measurement of SwissFEL, the knowledge of the gas jet diameter determines the maximum interaction length, which is used to assess the Guoy phase shift of the Terahertz wave \cite{THzSCModeling}.

The interaction point of the X-ray pulse and the exit nozzle of the gas jet is given by the size of the ion or electron spectrometers, respectively. We have thus set up a measurement of a piezo gas jet (Amsterdam Piezovalve) with an opening time of \SI{7}{\micro\second}. 
We present here measurements performed \SI{30}{\milli\metre} away from the nozzle.\ In order to measure the gas density up to the relevant distance the piezo valve is mounted on a linear vacuum stage.\ The opening time of the valve is set to \SI{55}{\micro\second} to measure the static gas flow without opening and closing effects of the valve (exposure time of CCD: \SI{30}{\micro\s}).\ Figure \ref{fig.piezo_radial} shows the radial density distribution, at distances of \SIrange{5}{29}{mm} from the piezo valve, at \SI{6}{\bar} backing pressure.\ In the same figure, the full width at half maximum of the radial distribution is shown.\ The gas expands linearly with a half opening angle of \SI{9}{\degree} in the considered region.\ Figure \ref{fig.piezo_cendens} depicts the central density of the gas flow.\ After expanding over a distance of almost \SI{30}{mm} the density is reduced by a factor of 20 down to \SI{1.5e15}{\per\cubic\centi\metre}. 

Similar methods may be used to study the electron beam that is used to generate the X-rays in a free electron laser. The tunnel ionization rate of the gas atoms depends on the peak current in the beam, which is a key parameter for the amplification in the FEL. 
The knowledge of the gas density, and its distribution, are important for the understanding of the data.

\subsection{Solenoid gas jet for LWFA}
A disadvantage of piezo valves is that the valve closing mechanism can only sustain backing pressures  $\leq\SI{10}{bar}$.\ Therefore, such valves are not suitable for high density applications.\ For the purpose of LWFA characterization, high-Z neutral atom densities are required, which are larger by one order of magnitude than the peak density, which the piezo valve can produce.\ The Parker solenoid valve (figure \ref{fig.valve}) is expected to provide densities exceeding \SI{1e18}{\per\cubic\centi\metre} at a distance of 2.6 mm from the nozzle, when operated with backing pressures around \SI{35}{bar} (figure \ref{fig.onaxisdensity}). 
\begin{figure}[h!]
    \centering
    \begin{subfigure}[b]{0.43\textwidth}
    \includegraphics[height=\textwidth]{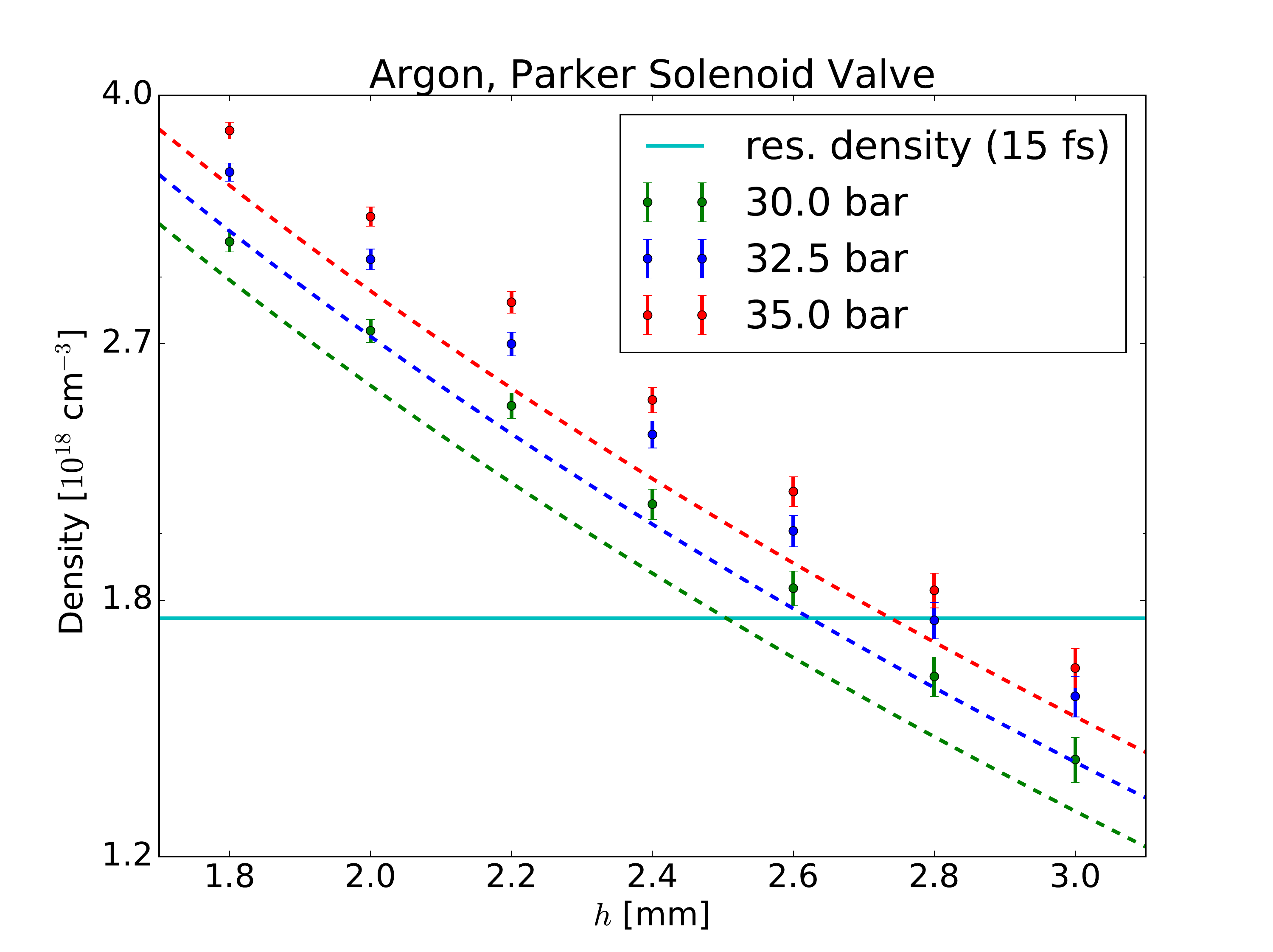}
    \end{subfigure} 
    \hfill
    \begin{subfigure}[b]{0.43\textwidth}
    \includegraphics[height=\textwidth]{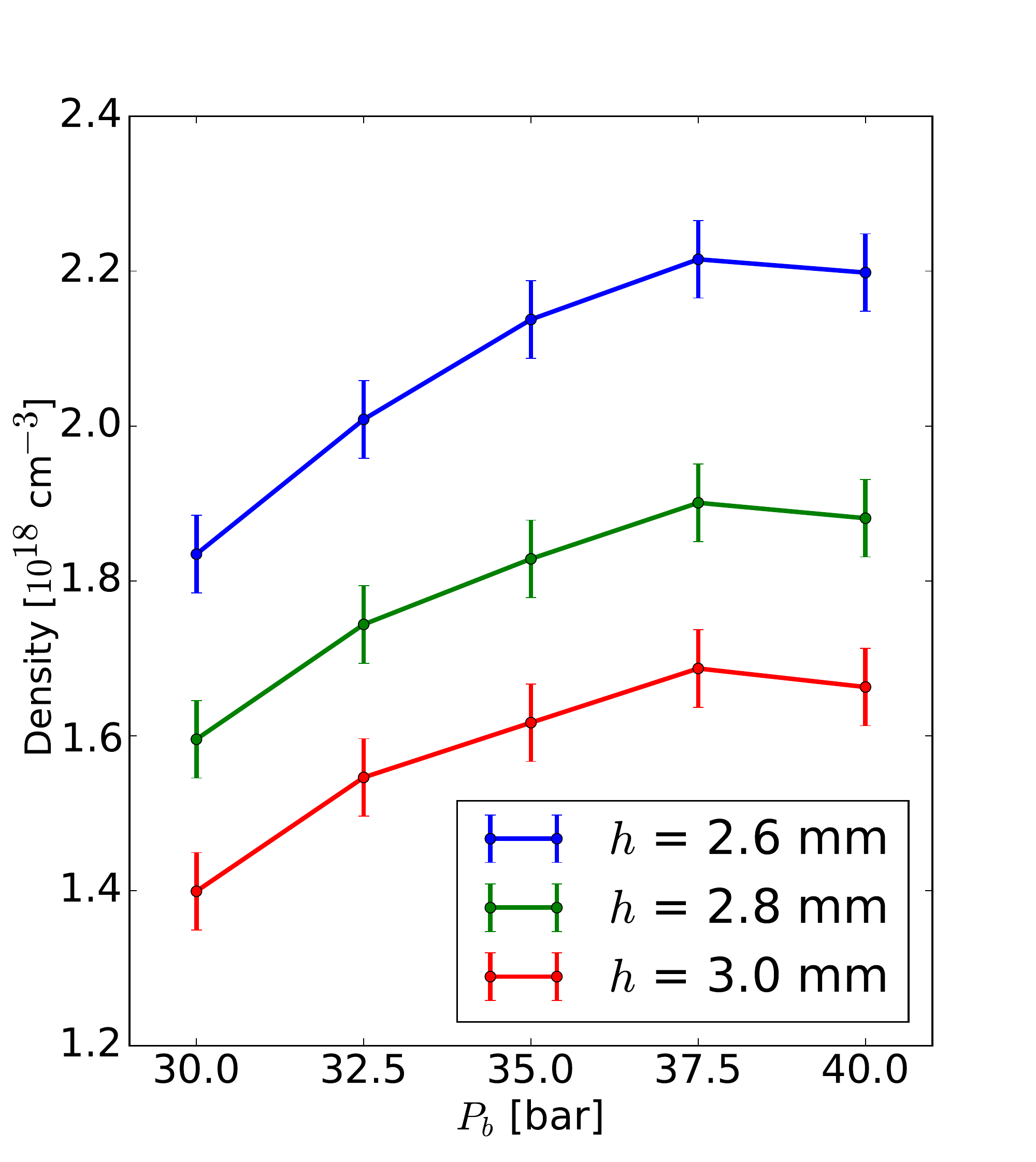}
    \end{subfigure}
    \caption{ML--EM reconstructed on-axis density of the solenoid valve without blade.\ Left: Dashed lines  represent the estimate for the respective backing pressure. The data points with error bars are from the measurements.\ The resonant density is reached with backing pressures between 30 and 35 bar in the region of interaction for the Ti:Sa pulse ($h=2.6-2.8$ mm).\ Right: The density appears to increase for pressures up to 37.5 bar.\ For higher pressures stagnation and even regression is observed.} \label{fig.solenoid_cendens}
\end{figure}
Due to geometric limitations in our set-up (blade holder) the focused laser beam in the LWFA set-up cannot interact with the gas closer to the nozzle than \SI{2.6}{\milli\metre}.\ The plasma frequency $\omega_p$, which is determined by the plasma density has to be matched to the laser pulse duration in order to excite the plasma wake efficiently.\ In the experiment the plasma frequency can be controlled by the density and ultimately by the backing pressure applied to the valve.\ The central density measurement of the solenoid gas jet (without blade) for argon is summarized in figure \ref{fig.solenoid_cendens}.\ 
From this plot one can infer the required backing pressure to create the resonant electron density at a certain distance from the nozzle.\ For a laser pulse of $\tau = \SI{15}{\femto\s}$ duration (FWHM) the required argon density is \SI{1.75e18}{\per\cubic\centi\metre}, assuming that the gas is ionised eight-fold (all valence electrons) \cite{nick}.\ Figure \ref{fig.solenoid_cendens} also contains the density estimates from section \ref{sec.est_dens}.\ Good qualitative agreement between the model and the measurement is observed.\ However, the measured densities are systematically higher than the estimate.\ A reason may be the fact that the estimate is based on the straight streamline model, which assumes that the expansion angle of the gas flow is equal to the opening angle of the nozzle \cite{chen}.\ However, the measured expansion angle, defined by a linear fit to the FWHM of the gas distribution, is around \SI{52}{\degree} which is significantly smaller than the opening angle of the conical nozzle (\SI{90}{\degree}).\ This would result in a higher particle density and could explain the discrepancy between the used model and the measurement results.\ Figure \ref{fig.solenoid_cendens} (right) depicts the argon density with respect to the backing pressure applied to the valve at fixed vertical distances $h$.\ It is observed that the argon density is increasing for backing pressures up to 37.5 bar.\ For even higher pressures the data indicates a stagnation and even a regression of the density.\ This can be explained by the working principle of the solenoid valve.\ When operating at higher backing pressures, a larger force is needed to lift the puppet out of the seal.\ Therefore, the valve may not open properly with too high backing pressures.\ This can then result in a smaller density.\ To overcome the force due to the backing pressure a high--voltage pulse (burst) is applied to the valve whose duration can be set internally.\ For the measurements shown here the burst duration was set to \SI{220}{\micro\s}.\

\subsection{Shock Front Characterisation in a LWFA}
\begin{figure}
    \centering
    \includegraphics[width=0.75\textwidth]{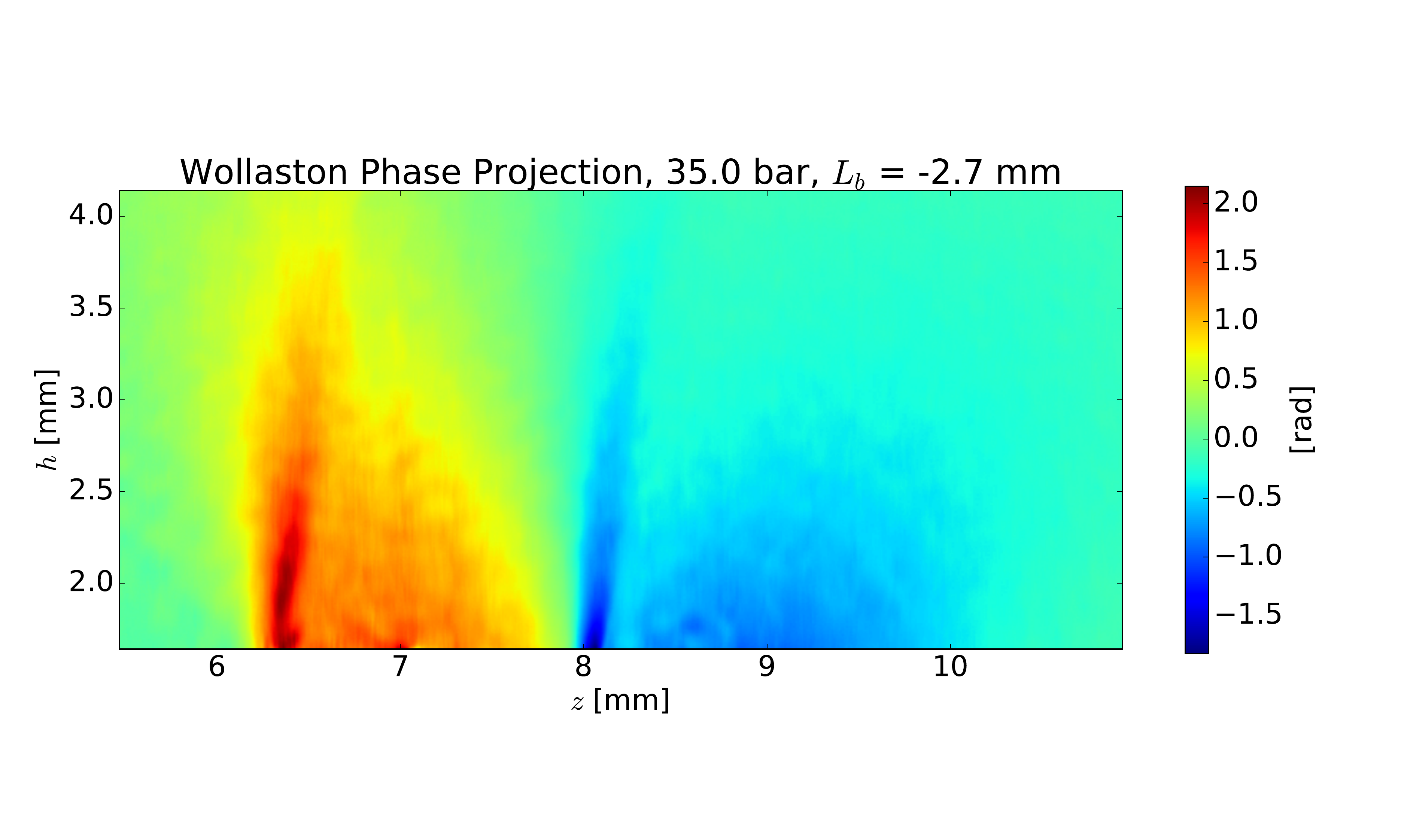}
    \caption{Typical Wollaston phase image of a shock front generated by a razor blade inserted from the left to a gas jet.\ Gas flow is directed upwards.}
    \label{fig.phase_shock} 
\end{figure}

\begin{figure}
    \centering
    \includegraphics[width=\textwidth]{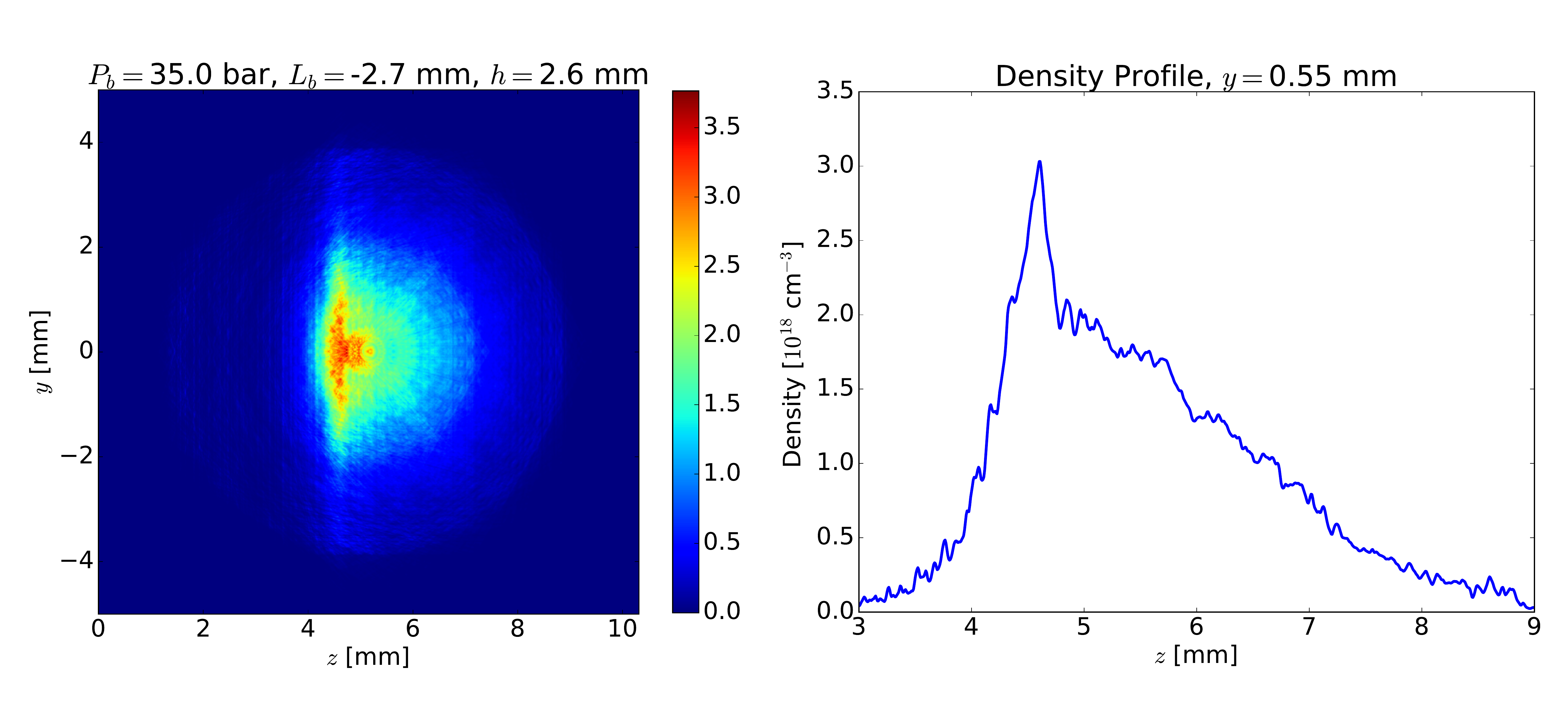}
    \caption{ML--EM reconstructed density distribution in a plane perpendicular to the gas flow at distance $h = \SI{2.6}{mm}$ from the nozzle, i.e.\ \SI{1}{mm} from the blade.}
    \label{fig.m_ramp_profile} 
\end{figure}

\begin{figure}
    \centering
    \includegraphics[width=0.5\textwidth]{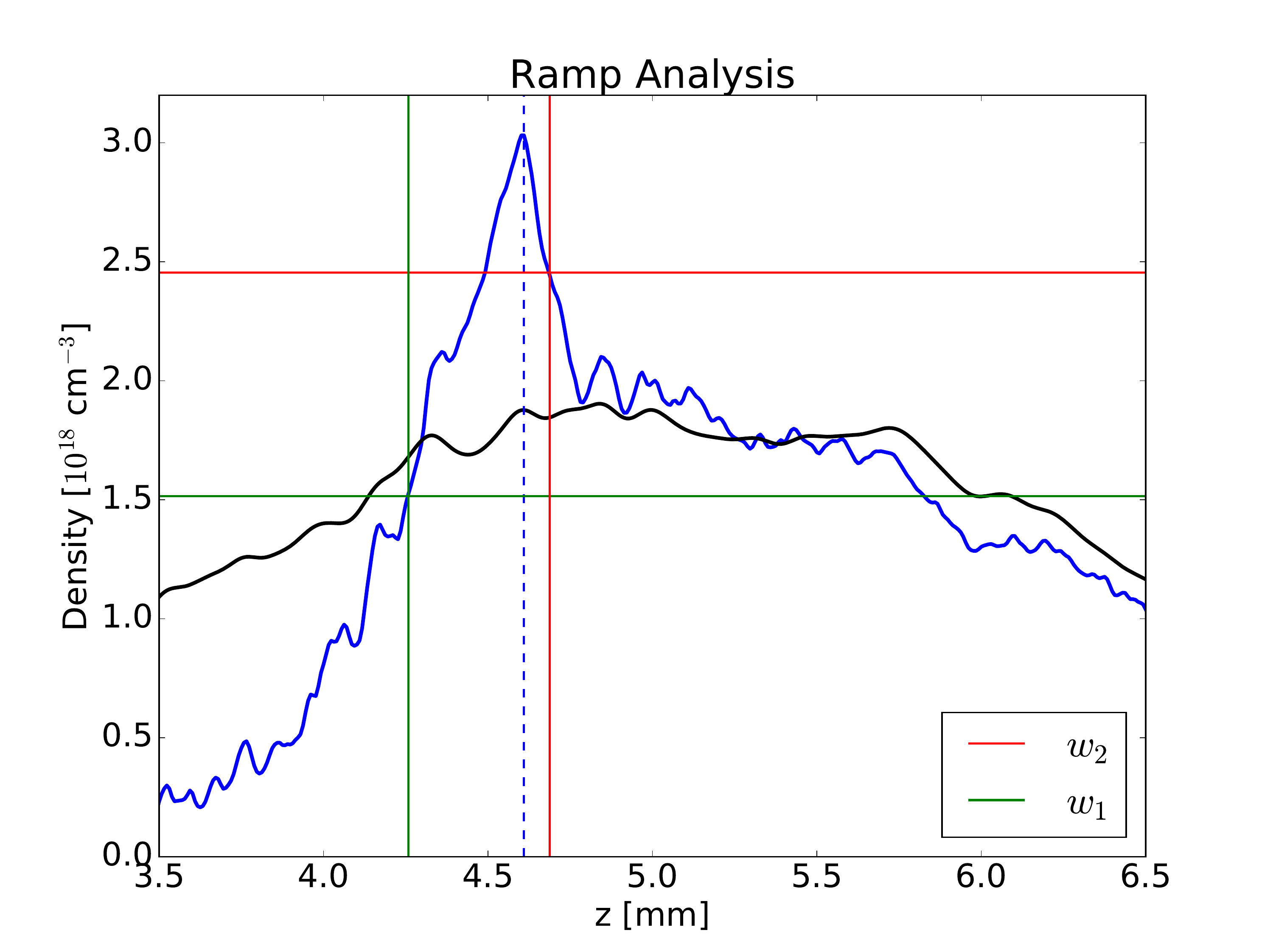}
    \caption{Example for ramp parameter analysis, undisturbed density in black.}
    \label{fig.ramp_ana_ex} 
\end{figure}

\begin{figure}
  \begin{subfigure}[b]{0.32\textwidth}
    \includegraphics[height=\textwidth]{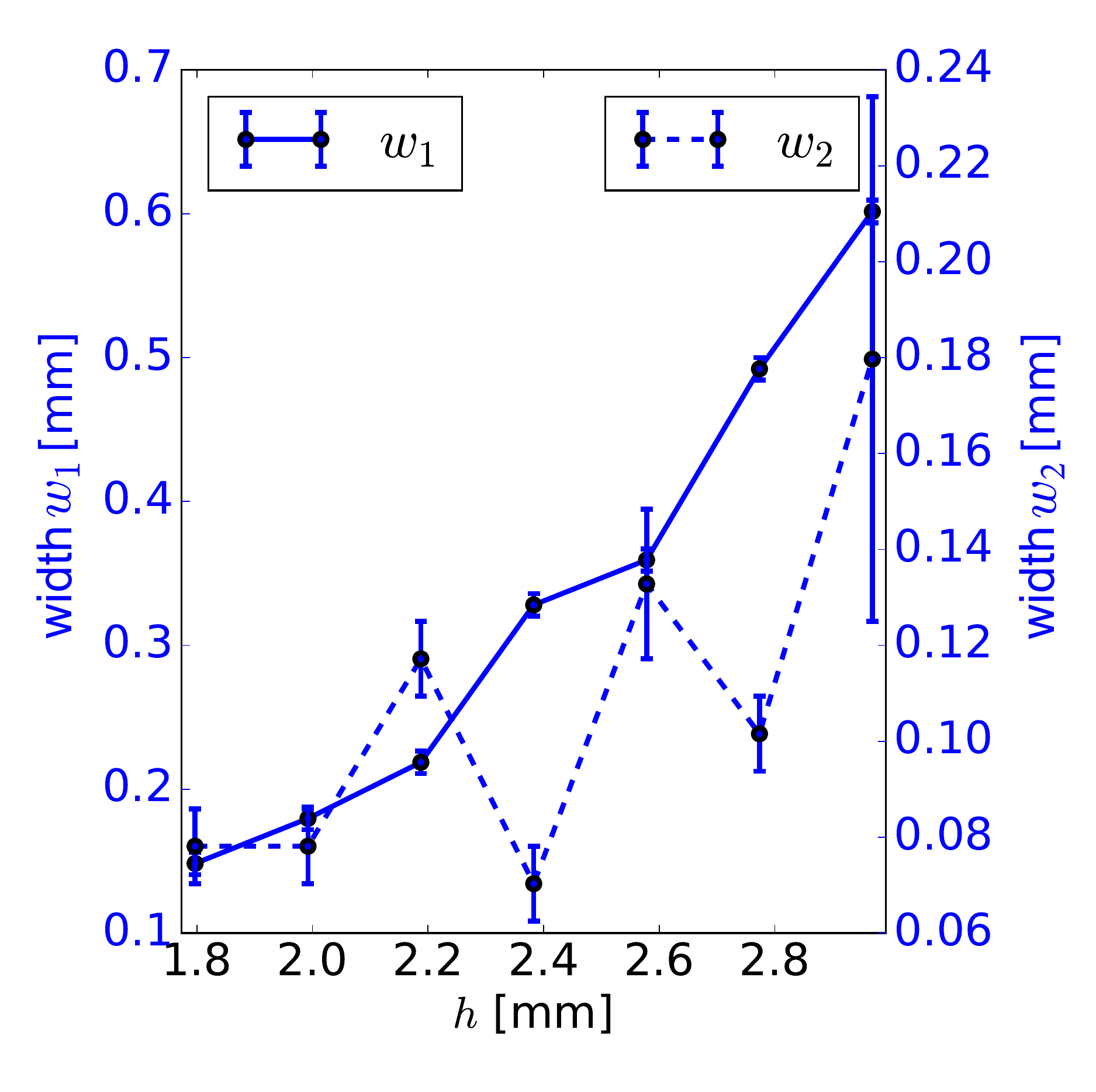}
  \end{subfigure}
    \begin{subfigure}[b]{0.32\textwidth}
    \includegraphics[height=\textwidth]{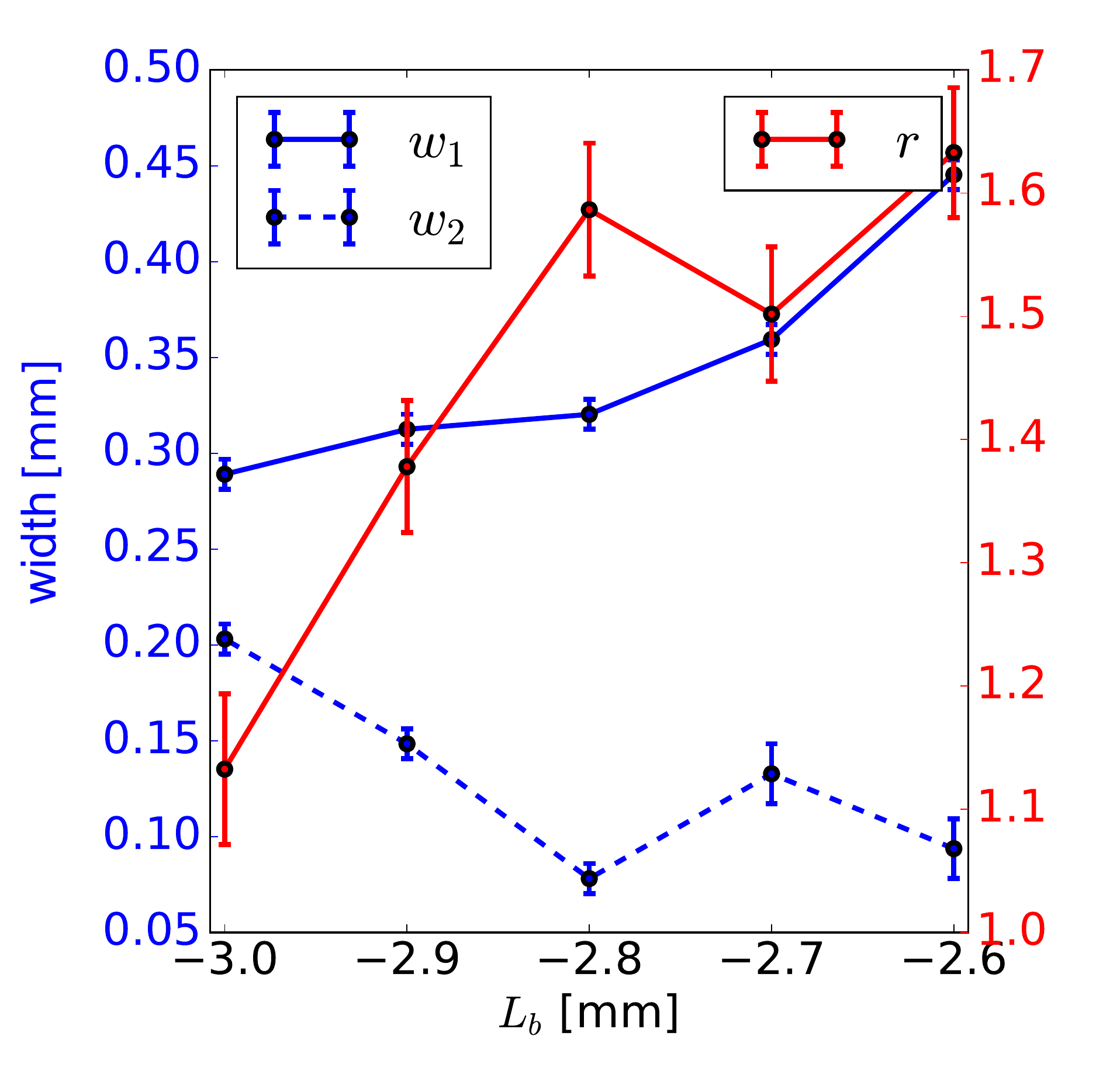}
  \end{subfigure}
    \begin{subfigure}[b]{0.32\textwidth}
    \includegraphics[height=\textwidth]{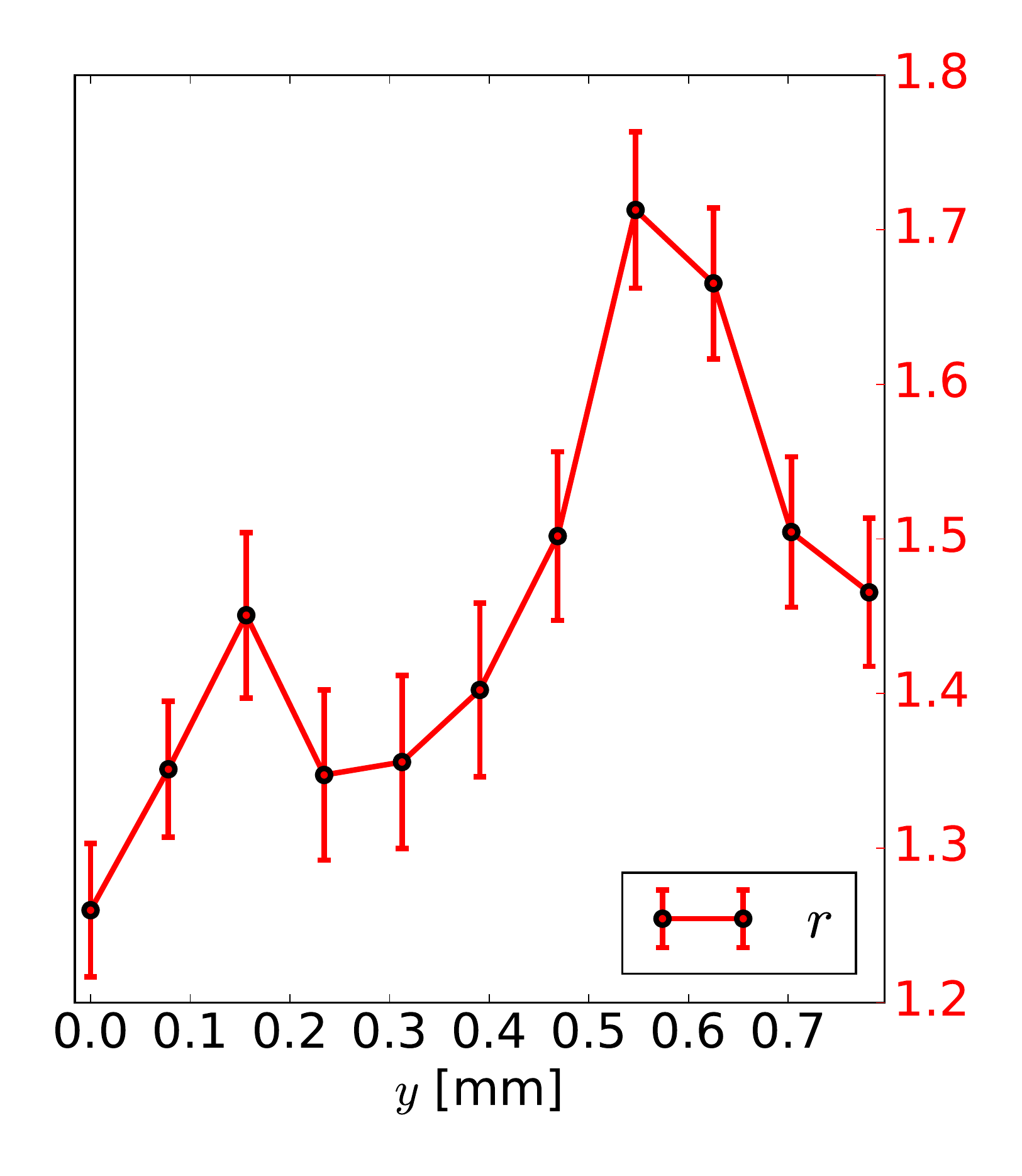}
  \end{subfigure}
  \caption{Ramp characterization with respect to $h$, $L_b$ and $y$.}\label{fig.ramp_ana}
\end{figure}

For the purpose of density down ramp injection in the linear regime of LWFA, a razor blade is inserted laterally into the gas jet to create a supersonic shock front \cite{PhysRevAccelBeams.20.051301}.\ A typical phase projection of the shock front is given in figure \ref{fig.phase_shock}.\ Two images next to each other are shown, one with positive and one with negative signal values. This is an artefact due to the working principle of the Wollaston prism, since it produces two parts in the interferograms, one represents the reference (unperturbed) and the other one the signal (c.f.\ section \ref{sec:theory}).\ From the projection it can already be noted that the density gradient is decreasing with increasing vertical distance above the nozzle.\ The tomographic data obtained with the rotational set-up, c.f.\ figure \ref{fig.tommodule}, is analysed via the ML--EM reconstruction algorithm explained in section \ref{sec.mlem}.\ Figure \ref{fig.m_ramp_profile} depicts the reconstructed density distribution of the shock front at 35 bar backing pressure.\ To study and optimise the ramp properties, this measurement is carried out for different backing pressures $P_b$ and blade positions $L_b$ from the center up to the edge of the gas flow (edge: $L_b=\SI{-3.2}{mm}$).\ The ML--EM tomography is computed at several distances from the nozzle $h$.\ The range of these parameters is summarised in Table \ref{tab.parameters}.\  In order to evaluate the characteristics of the shock front numerically, the following quantities are defined:
\begin{itemize}
    \item Height $h_s$ of the shock front: Density difference between ramped and undisturbed distribution at the ramp,
    \item Ramp factor $r$: Ramped peak density divided by undisturbed peak density,
    \item $w_1$: Half--width (left) defined by the ramp peak density,
    \item $w_2$: Half--width (right) defined by the height $h_s$.
\end{itemize}
For a better understanding, the parameters $w_1$ and $w_2$ are indicated in figure \ref{fig.ramp_ana_ex}, showing the density of a shock front as well as the undisturbed distribution along the $z$--direction.\ The main effect of the backing pressure is the central density which is plotted in figure \ref{fig.solenoid_cendens}.\ The ramp characteristics are governed by the parameters $L_b$, $h$ and $y$ and their respective effects are summarised in figure \ref{fig.ramp_ana} at a fixed backing pressure of 35 bar.\ A sharp density down--ramp is desirable to achieve a small electron energy spread in a LWFA (\cite{PhysRevLett.100.215004}, \cite{Schmid}, \cite{Buck},  \cite{He2013}) .\ The length of the down--ramp is quantified by $w_2$ which is increasing with distance from the blade.\ A slight dependency of $w_1$ on the blade position $L_b$ is observed with a local minimum at $L_b$ = -2.8 mm.\ Another important quantity is the ramp factor $r$, which is mainly determined by $L_b$ and $y$.\ It is larger when the blade is positioned closer to the center of the gas flow, as the ramp is created in a region of higher undisturbed density.\ The measurements indicate thact the ramp factor also depends on the horizontal position $y$.\ In particular, $r$ has a local maximum 0.5-0.6 mm displaced from the center.\ This is understandable by looking at profiles of a two dimensional Gaussian.\ When the profile is off--centered the distribution is flatter.\ This means that for a centered profile the density is rising more after the ramp,  which results in a lower ramp factor.\

\begin{table}
\centering
\caption{Parameters for shock front measurements.}
\label{tab.parameters}
\begin{tabular}{c|c|c|c|c}
            & $P_b$ [bar]   & $L_b$ [mm]    & $h$ [mm]& $h$ [pixel] \\ \cline{1-5}
min, max    & 30.0, 40.0    & -3.2, -2.5    & 1.8, 3.0  & 1200, 1050 \\ \cline{1-5}
step        & 2.5           &  0.1          & 0.2 & 25 
\end{tabular}
\end{table}

\subsection{Error and Stability Analysis}
For the phase measurements of the piezo valve an average of 1000 images is taken in order to reduce noise.\ This leads to an error of below \SI{1}{\milli\radian} (standard deviation $\sigma$), which is needed in order to measure low densities down to \SI{1.5e15}{\per\cubic\centi\metre} with an uncertainty of \SI{0.4e15}{\per\cubic\centi\metre}.\ 

\begin{figure}
    \centering
    \includegraphics[width=\textwidth]{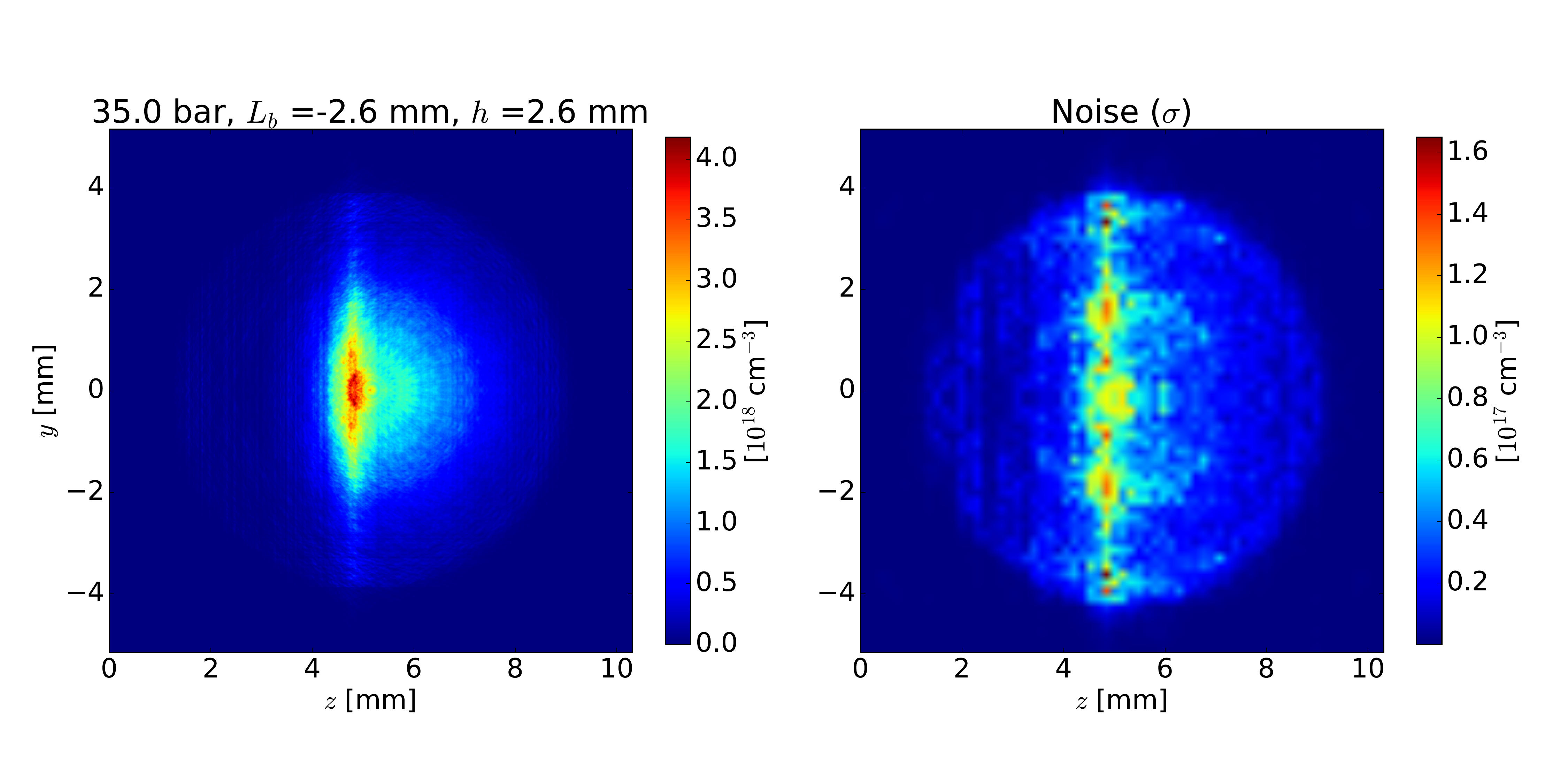}
    \caption{Error analysis of the reconstructed density (ML--EM, 15 iterations).}
    \label{fig.mlem_dens_error} 
\end{figure}

In case of the solenoid valve, the purpose is to create and measure a shock front with densities up to 3 magnitudes larger than in the scenario described above.\ Combined with the fact that tomography requires more data (phase projections from different angles), it is decided to reduce the number of images per measurement drastically.\ For instance, the peak density in the shock front at 35 bar, 1 mm away from the nozzle is 1.5 rad and has a standard deviation of \SI{0.05}{\radian}.\ After averaging, the phase signal has noise in the order of \SI{0.01}{\radian}.\ As the maximum of the phase signal is around \SI{1}{\radian}, the ML--EM test (section \ref{sec.mlem_test}) is realistic with a noise level of $\sigma=$ \SI{0.01}{}.\ (The generic distribution for this test is normalized, such that the maximum of the projections is equal to 1.) In agreement with the ML--EM convergence test, the number of iterations for the 3D density reconstruction is set to 10. Figure \ref{fig.mlem_dens_error} shows the density distribution of the shock front at 35 bar after 10 iterations and the noise of the reconstructed image.\ The noise is calculated as follows.\ The data is fitted with a Savitzgy--Golay filter, which smoothes noisy data by interpolation within a symmetric window around each data point (window size: 51 pixel, polynomial order: 3) \cite{sav-gol,LUO20051429}.\ The fit is subtracted from the noisy data, then divided into segments of 30 pixel (0.2 mm), over which the standard deviation is calculated.\ This provides a local error estimate of the reconstructed density.\ The noise ($\sigma < \SI{1.4e17}{\per\cubic\centi\metre}$) in the region of the shock front is a factor of 20 lower than the peak density \SI{2.8e18}{\per\cubic\centi\metre}, i.\ e.\ the density change in the shock front is reconstructed with sufficient precision.

\section{Conclusion} \label{sec:Conclusion}
A simple single beam Wollaston interferometer has been described. The interferometer was used to measure gas densities in electron beam monitors for the free electron laser and to characterise shock fronts in a LWFA. Convergence and error studies of the used ML--EM algorithm show adequate accuracy of the presented problems. The use of parallel computation and GPU technology reduces the computational time below the data taking time. In that sense this represents real time density computation.

The presented set-up is limited by the slow rotational stage and the low frequency of the gas jet
due to the weak vacuum pump. These technical limits can be easily overcome, hence the time for a full 3D density reconstruction is estimated to be less than \SI{10}{\second}. 

We plan to explore the possibility of interleaving the measurement with the operation of the LWFA.  We aim at periodic
measurements of the gas-jet density and potentially integrate the obtained results, in a feedback
loop, with the overall goal to improve quality and stability of the electron beam.

Further studies will also include the investigation of better initial conditions w.r.t. the convergence of the ML--EM algorithm and the understanding of the artefacts such as the small peaks at the centre and the concentric rings around this point, visible in 
the 2D density reconstructions.

\section*{Acknowledgements}
We wish to thank the following individuals for their support: Christoph Hauri, Yunpei Deng, Christian Erny, Julien R\'{e}hault, Peter Radi, Ivo Alxneit, Martin Beck, Franziska Frei, Martin Rohrer, Hanspeter Gehrig and Dominique Zehnder.
\bibliographystyle{elsarticle-num}  
\bibliography{refs}{}
\end{document}